\documentclass[prx,twocolumn,showpacs,preprintnumbers,amsmath,amssymb]{revtex4-1}
\usepackage{graphicx}
\usepackage{amsmath}
\usepackage{MnSymbol}
\usepackage{mathrsfs}
\usepackage{colortbl}
\usepackage[table]{xcolor}
\usepackage{tabularx}
\makeatletter
\newcommand{\boldnabla}{\mbox{\boldmath$\nabla$}}
\newcommand{\boldmu}{\mbox{\boldmath$\mu$}}

\makeatother

\begin{document}

\title{Chiral Rotational Spectroscopy}
\author{Robert P. Cameron} 
\email{robert.cameron@glasgow.ac.uk}
\affiliation{Max Planck Institute for the Physics of Complex Systems, Dresden D-01187, Germany}
\affiliation{School of Physics and Astronomy, University of Glasgow, Glasgow G12 8QQ, U.K.}
\author{J\"{o}rg B. G\"{o}tte}
\affiliation{Max Planck Institute for the Physics of Complex Systems, Dresden D-01187, Germany}
\affiliation{School of Physics and Astronomy, University of Glasgow, Glasgow G12 8QQ, U.K.}
\author{Stephen M. Barnett}
\affiliation{School of Physics and Astronomy, University of Glasgow, Glasgow G12 8QQ, U.K.}
\date{\today}
\begin{abstract}
We introduce chiral rotational spectroscopy: a new technique that enables the determination of the orientated optical activity pseudotensor components $B_{XX}$, $B_{YY}$ and $B_{ZZ}$ of chiral molecules, in a manner that reveals the enantiomeric constitution of a sample and provides an incisive signal even for a racemate. Chiral rotational spectroscopy could find particular use in the analysis of molecules that are chiral solely by virtue of their isotopic constitution and molecules with multiple chiral centres. A basic design for a chiral rotational spectrometer together with a model of its functionality is given. Our proposed technique offers the more familiar polarisability components $\alpha_{XX}$, $\alpha_{YY}$ and $\alpha_{ZZ}$ as by-products, which could see it find use even for achiral molecules.
\end{abstract}
\maketitle


\section{Introduction} 
\label{Introduction}
Chirality pervades the natural world and is of particular importance to life, as the molecules that comprise living things are chiral and their chirality is crucial to their biological function \cite{F1, Lough 02, Barron 04, Blackmond 10}. Our ability to probe and harness molecular chirality remains incomplete in many respects, however, and new techniques for chiral molecules are, therefore, highly sought after. 

The basic property of a chiral molecule that is probed in typical optical rotation experiments using fluid samples \cite{Barron 04, Biot 15, Rosenfeld 28, Craig 98, Atkins 11} is the isotropic sum
\begin{equation}
{\textstyle\frac{1}{3}}(B_{XX}+B_{YY}+B_{ZZ})
\label{isotropic sum}
\end{equation}
with $B_{XX}$, $B_{YY}$ and $B_{ZZ}$ components of the optical activity pseudotensor \cite{Buckingham 71, Autschbach 11} referred to molecule-fixed axes $X$, $Y$ and $Z$. These experiments yield no information about $B_{XX}$, $B_{YY}$ or $B_{ZZ}$ individually. Other well-established chiroptical techniques such as circular dichroism \cite{Lough 02, Barron 04, Craig 98, Cotton 95 a, Cotton 95 b, Holzwarth 74, Barron 10} and Raman optical activity \cite{Barron 04, Craig 98, Barron 10, Atkins 69, Barron 71, Barron 73, Barron 07} yield other chirally sensitive molecular properties but the fact remains that it is the isotropically averaged forms of these that are usually observed in practice.

The ability to determine orientated rather than isotropically averaged chiroptical information, in particular the individual, orientated components $B_{XX}$, $B_{YY}$ and $B_{ZZ}$ \cite{Caveatt} is highly attractive, as these offer a wealth of information about molecular chirality that is only partially embodied by the isotropic sum (\ref{isotropic sum}). At present such information can only be obtained, however, using an orientated sample as in a crystalline phase \cite{Kaminsky 00, Kahr 12}. The preparation of such samples is not always feasible and even when it can be achieved, signatures of chirality are usually very difficult to distinguish from other effects, in particular those due to linear birefringence. Indeed, it was noted in 2012 that ``\textit{we have a shockingly small amount of data on the chiroptical responses of orientated molecules, a vast chasm in the science of molecular chirality}" \cite{Kahr 12}.
 
In the present paper we introduce chiral rotational spectroscopy: a new technique with the ability to
\begin{enumerate}
\item[\textbf{(i)}] determine $B_{XX}$, $B_{YY}$ and $B_{ZZ}$ individually, thus promising to fill the ``\textit{vast chasm}'' described above;
\item[\textbf{(ii)}] measure the enantiomeric excess of a sample and provide an incisive signal even for a racemate, thus negating the need for dissymmetric synthesis or resolution, which are instead required by traditional techniques;
\item[\textbf{(iii)}] probe the chirality of molecules that are chiral solely by virtue of their isotopic constitution, the importance of which is becoming increasingly apparent whilst traditional techniques remain somewhat lacking in their sensitivities;
\item[\textbf{(iv)}] distinguish clearly and in a chirally sensitive manner between subtly different molecular forms, making it particularly useful for molecules with multiple chiral centres, the analysis of which using traditional techniques represents a serious challenge.
\end{enumerate}
Chiral rotational spectroscopy is distinct from another class of techniques introduced recently, in which the phase of a microwave signal is used to discriminate between opposite enantiomers \cite{Hirota 12, Nafie 13, Patterson 13, Patterson 13b, Shubert 14a, Shubert 14, Lobsiger 14, Lehman 15a, Shubert 15, Shubert 15 b}. We refer to these collectively as `chiral microwave three wave mixing'. In what follows we will compare and contrast chiral rotational spectroscopy and chiral microwave three wave mixing. It is our hope that these techniques will one day complement each other.

Let us emphasise that chiral rotational spectroscopy boasts the abilities \textbf{(i)}-\textbf{(iv)} \textit{simultaneously}. Various other techniques boast \textit{some} of these abilities but none offer them \textit{all}, together. For example: vibrational circular dichroism and Raman optical activity are inherently sensitive to isotopic molecular chirality to leading order as per \textbf{(iii)} \cite{Holzwarth 74, Barron 77, Barron 78} but are blind to racemic samples and so cannot fully realise \textbf{(ii)}, nor are they particularly well suited for \textbf{(iv)}; chiral microwave three-wave mixing can realise \textbf{(iv)} \cite{Shubert 14a, Shubert 14, Shubert 15, Shubert 15 b} but is also blind to racemic samples and so cannot fully realise \textbf{(ii)}; Coloumb explosion imaging can probe the molecular chirality of racemates as per \textbf{(ii)} \cite{Kitamura 01, Pitzer 13} and is inherently sensitive to isotopic molecular chirality to leading order as per \textbf{(iii)} \cite{ Pitzer 13}, but is seemingly restricted in its application to relatively simple structures under well-controlled conditions and is not particularly well suited for realising \textbf{(iv)}.

We work in a laboratory frame of reference $x$, $y$ and $z$ with time $t$ and $\hat{\mathbf{x}}$, $\hat{\mathbf{y}}$ and $\hat{\mathbf{z}}$ unit vectors in the $+x$, $+y$ and $+z$ directions. A lower-case index $a$ can take on the values $x$, $y$ or $z$ and upper-case indices $A$, $B$ and $C$ can take on the values $X$, $Y$ or $Z$. The summation convention is to be understood throughout.


\section{Chiral rotational spectroscopy}
\label{Chiralrotationalspectroscopy}
In the present section we elucidate the basic premise of chiral rotational spectroscopy: \textit{chiral molecules illuminated by circularly polarised light yield orientated chiroptical information via their rotational spectrum}. 

To enable the determination of orientated chiroptical information we recognise the need to
\begin{enumerate}
\item[\textbf{(i)}] prepare a chiral sample of orientated character, 
\item[\textbf{(ii)}] evoke a chiroptical response from the sample and
\item[\textbf{(iii)}] observe and interpret the response so as to obtain orientated chiroptical information.
\end{enumerate}
\noindent We envisage fulfilling these objectives as follows.

\begin{figure}[h!]
\centering
\includegraphics[width=0.6\linewidth]{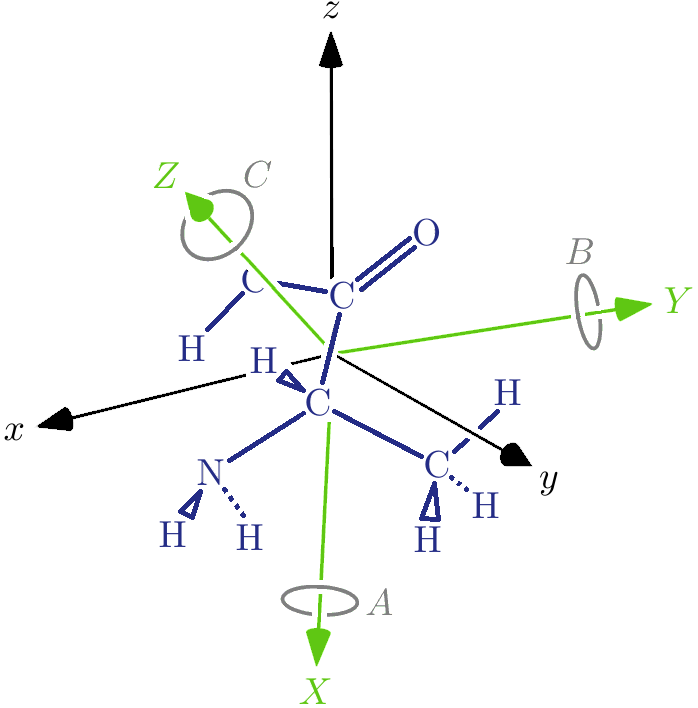}\\
\caption{\small An un-ionised \textsc{L}-$\alpha$-alanine molecule. \textsc{L}-$\alpha$-alanine is an amino acid found in abundance in living things \cite{Lough 02, Bunker 05, Blackmond 10}. Produced using data from \cite{CCC BDB}.} 
\label{EulerDiagram}
\end{figure}

\subsection{Chiral sample of orientated character}
\begin{figure}[h!]
\centering
\includegraphics[width=1\linewidth]{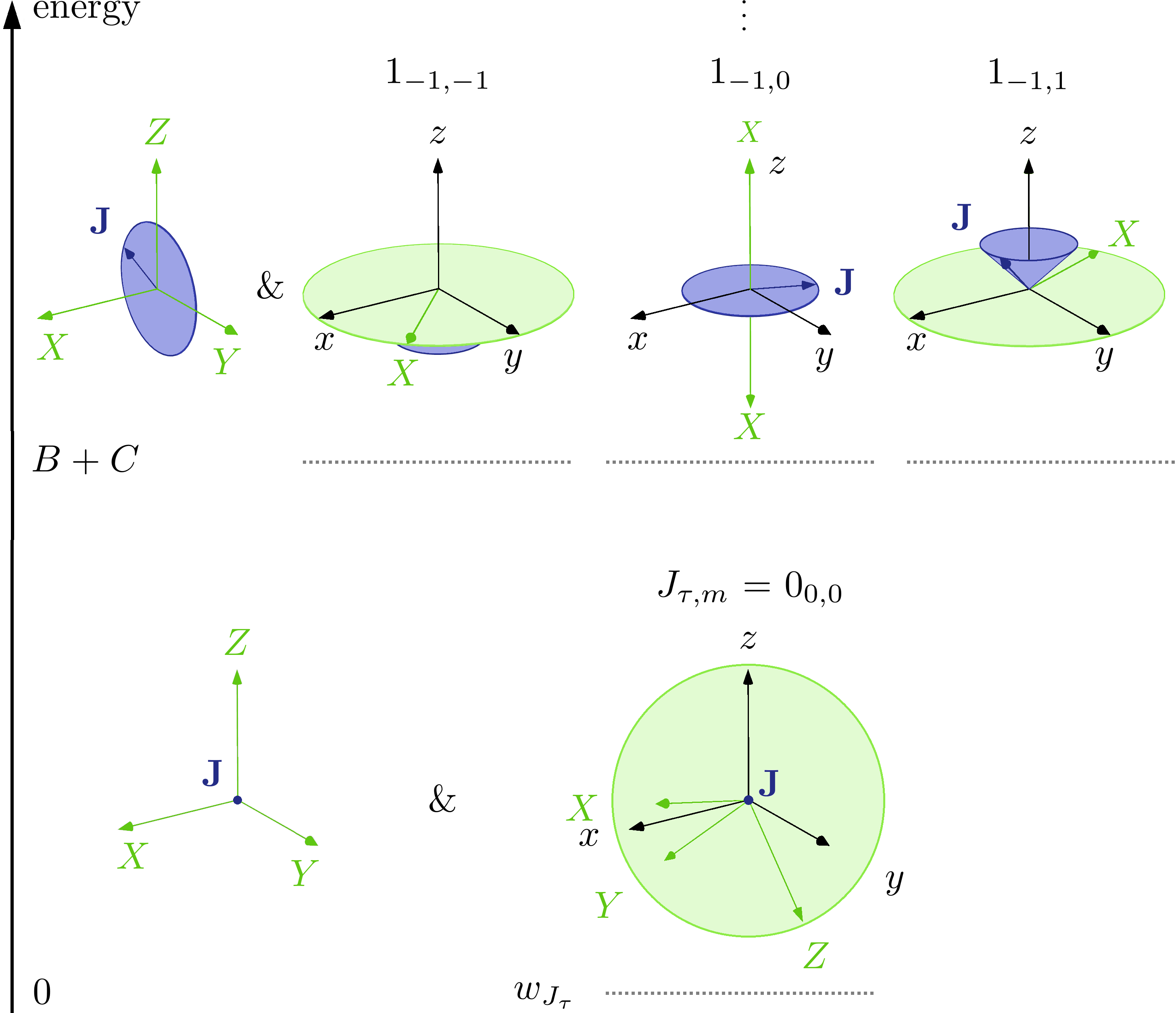}\\
\caption{\small The blue regions here depict equally probable values of the rotor angular momentum $\mathbf{J}$ relative to the molecule-fixed axes $X$, $Y$ and $Z$ and to the laboratory-fixed axes $x$, $y$ and $z$ for some of the molecule's rotor states whilst the green regions indicate the most probable orientations of $X$, $Y$ and $Z$ relative to $x$, $y$ and $z$.} 
\label{Rotationgeometry}
\end{figure}

A chiral molecule with unimpeded rotational degrees of freedom, as in the gas phase or a molecular beam \cite{Townes 55, Ramsey 56}, can already be regarded as a chiral sample of orientated character, as we will now demonstrate. Let us assume that the molecule is at rest or moving slowly and that it occupies its vibronic ground state, in which it is small, polar and non-paramagnetic \cite{Born 27, Brown 03, Atkins 11, Bunker 05, Bernath 05}. We model the rotation of the molecule as that of an asymmetric rigid rotor, with equilibrium rotational constants $A>B>C$ associated with rotations about the molecule-fixed, principal axes of inertia $X$, $Y$ and $Z$, as depicted in FIG. \ref{EulerDiagram}. The rotational and nuclear-spin degrees of freedom of the molecule should be well described \cite{F3} then by the effective Hamiltonian
\begin{equation}
\hat{H}=\hat{H}_\textrm{rotor}+\delta\hat{H}
\end{equation}
with 
\begin{equation}
\hat{H}_\textrm{rotor}=\frac{1}{\hbar^2}\left(A\hat{J}_X^2+B\hat{J}_Y^2+C\hat{J}_Z^2\right)
\end{equation}
the rotor Hamiltonian \cite{Wang 29, Townes 55, Bunker 05, Bernath 05} and $\delta\hat{H}$ accounting for nuclear spin \cite{Kellog 39, Bragg 48, Bragg 49, Townes 55, White 55, Ramsey 56, Flygare 74, Brown 03} and perhaps also corrections to the rigid rotor model such as those due to centrifugal distortion \cite{Townes 55, Bunker 05, Atkins 11}. The components $\hat{J}_A$ of the rotor angular momentum account for the entirety of the molecule's intrinsic angular momentum except for nuclear spin \cite{Bunker 05}. Let us neglect $\delta\hat{H}$ for the moment and focus our attention upon the rotor states $|J_{\tau,m}\rangle$ and rotor energies $w_{J_{\tau}}$, which satisfy
\begin{equation}
\hat{H}_\textrm{rotor}|J_{\tau,m}\rangle=w_{J_{\tau}}|J_{\tau,m}\rangle
\end{equation}
with $J\in\{0,1,\dots\}$ determining the magnitude of the rotor angular momentum, $\tau\in\{0,\dots,\pm J\}$ labeling the rotor energy and $m\in\{0,\dots,\pm J\}$ determining the $z$ component of the rotor angular momentum \cite{Wang 29, Townes 55, Bunker 05, Bernath 05}. Some of these are depicted in FIG. \ref{Rotationgeometry}. In the $J_{\tau,m}=0_{0,0}$ rotor state the molecule possesses a vanishing rotor energy of $w_{0_0}=0$, as it is not rotating. All orientations of $X$, $Y$ and $Z$ relative to $x$, $y$ and $z$ are, therefore, equally likely to be found. In the $1_{-1,m}$ rotor states, however, the molecule possesses a rotor energy of $w_{1_{-1}}=B+C$, as it will never be found rotating about the $X$ axis but is equally likely to be found rotating about the $Y$ or $Z$ axis. The conceivable motions of the rotor then conspire such that for $m=0$ the $X$ axis is most likely to be found perpendicular to the $x$-$y$ plane whereas for $m=\pm1$ it is most likely to be found in the $x$-$y$ plane. Analogous observations hold for the $1_{0,m}$ rotor states, in which it is the $Y$ axis that is treated preferentially, and the $1_{1,m}$ rotor states, in which it is the $Z$ axis. They can be extended, moreover, to the $J\in\{2,\dots\}$ manifolds, although the analysis becomes increasingly complicated with increasing $J$. The important point here is that the rotation and hence orientation of the molecule in any particular rotor state is \textit{not} of isotropic character, in general. Indeed, the most probable orientations of the molecule differ for different rotor states. The isotropic character usually ascribed to the gas phase or a molecular beam emerges only when these states are suitably averaged over, in accord with the principal of spectroscopic stability \cite{Vleck 32}. Such observations are complicated by the inclusion of $\delta\hat{H}$, but only superficially. 

\subsection{Evoking a chiroptical response from the sample}
Suppose now that the molecule is illuminated by far off-resonance visible or perhaps near infrared circularly polarised light of moderate intensity $I$ and wavevector $\mathbf{k}$ pointing in the $z$ direction, with the ellipticity parameter $-1\le\sigma\le1$ of the light taking its limiting values here of $\pm1$ for left- or right-handed circular polarisation \cite{Newfootnote}. The light simply \cite{F6} drives oscillations in the charge and current distributions of the molecule, biasing the rotation of the molecule whilst shifting its energy in an orientationally and chirally sensitive manner, as we will now demonstrate. Such shifts constitute our orientated chiroptical response. We extend well-established methods \cite{Flygare 74} by dressing the interaction between the light and the molecule's electrons \cite{F4} and find that the rotational and nuclear-spin degrees of freedom of the molecule should be well described by the effective Hamiltonian
\begin{equation}
\hat{H}'=\hat{H}_\textrm{rotor}+\hat{H}_\textrm{light}+\delta\hat{H}
\end{equation}
with
\begin{eqnarray}
\hat{H}_\textrm{light}&=&-\frac{I}{2\epsilon_0 c}\left(\hat{\ell}_{xA}\hat{\ell}_{xB}+\hat{\ell}_{yA}\hat{\ell}_{yB}\right) \\
& \times &\left(\frac{1}{2}\alpha_{AB}+\sigma |\mathbf{k}|B_{AB}\right) \nonumber
\end{eqnarray}
accounting for the energy associated with the oscillations \cite{Cameron 14 c}. The $\hat{\ell}_{aA}$ are direction cosines \cite{Barron 04, Bunker 05}, which quantify the orientation of the molecule relative to the light. The $\alpha_{AB}$ are components of the electronic electric-dipole / electric-dipole polarisability \cite{Vleck 32, Barron 04, Craig 98}, which quantify the susceptibility of the charge and current distributions of the molecule to be distorted by the light in a chirally insensitive manner: $\alpha_{XX}$, $\alpha_{YY}$ and $\alpha_{ZZ}$ in particular are identical for opposite enantiomers. The $B_{AB}$ are components of the electronic optical activity pseudotensor \cite{Buckingham 71, Autschbach 11}, which quantify the susceptibility of the charge and current distributions of the molecule to be distorted in a chirally sensitive manner: $B_{XX}$, $B_{YY}$ and $B_{ZZ}$ in particular each possess equal magnitudes but opposite signs for opposite enantiomers and are the molecular properties upon which chiral rotational spectroscopy is based. Let us focus our attention now upon a molecule with nuclear spins of $0$ or $1/2$ only and assume that $\hat{H}_\textrm{rotor}\gg\hat{H}_\textrm{light}\gg\delta \hat{H}$ with no accidental degeneracies of importance whilst neglecting the possibility of any effects due to the spin statistics of similar nuclei \cite{Townes 55, Ramsey 56, Bunkers 98, Atkins 11}. The energy of the perturbed $0_{0,0}$ rotor state together with a nuclear-spin state $|n\rangle$ is then essentially 
\begin{equation}
w_{0_0}+\Delta w_{0_{0,0}}+ \delta w_{0_{0,0}} 
\label{G12a}
\end{equation}
with
\begin{eqnarray}
w_{0_0}&=&0, \nonumber \\
\Delta w_{0_{0,0}}&=&\langle 0_{0,0}|\hat{H}_\textrm{light}|0_{0,0}\rangle+\dots \\
&=&-\frac{I}{\epsilon_0 c}\Bigg[\frac{1}{3}\left(\frac{1}{2}\alpha_{XX}+\sigma|\mathbf{k}|B_{XX}\right) \nonumber \\
&&+\frac{1}{3}\left(\frac{1}{2}\alpha_{YY}+\sigma|\mathbf{k}|B_{YY}\right) \nonumber \\
&&+\frac{1}{3}\left(\frac{1}{2}\alpha_{ZZ}+\sigma|\mathbf{k}|B_{ZZ}\right)\Bigg]+\dots \nonumber \\
\delta w_{0_{0,0}}&=&\langle n|\langle 0_{0,0}|\delta\hat{H}|0_{0,0}\rangle|n\rangle
\end{eqnarray}
the unperturbed rotor energy, an energy shift due to the light and a further energy shift due to nuclear intramolecular interactions and perhaps also corrections to the rigid rotor model, where we assume $|n\rangle$ to be diagonal in $\langle 0_{0,0}|\delta\hat{H}|0_{0,0} \rangle$. The components $\alpha_{XX}$, $\alpha_{YY}$, $\alpha_{ZZ}$, $B_{XX}$, $B_{YY}$ and $B_{ZZ}$ make isotropically weighted contributions, reflecting the idea that all orientations of the molecule relative to the light are equally likely to be found in the $0_{0,0}$ rotor state: the electric and magnetic field vectors of the light can be said to drive oscillations equally along the $X$, $Y$ and $Z$ axes. In contrast, the energy of the perturbed $1_{-1,0}$ rotor state together with a nuclear-spin state $|n'\rangle$ is essentially
\begin{equation}
w_{1_{-1}}+\Delta w_{1_{-1,0}}+\delta w_{1_{-1,0}} 
\label{G12b}
\end{equation}
with 
\begin{eqnarray}
w_{1_{-1}}&=&B+C, \nonumber \\
\Delta w_{1_{-1,0}}&=&\langle 1_{-1,0}|\hat{H}_\textrm{light}|1_{-1,0}\rangle+\dots \\
&=&-\frac{I}{\epsilon_0 c}\Bigg[\frac{1}{5}\left(\frac{1}{2}\alpha_{XX}+\sigma|\mathbf{k}|B_{XX}\right) \nonumber \\
&&+\frac{2}{5}\left(\frac{1}{2}\alpha_{YY}+\sigma|\mathbf{k}|B_{YY}\right) \nonumber \\
&&+\frac{2}{5}\left(\frac{1}{2}\alpha_{ZZ}+\sigma|\mathbf{k}|B_{ZZ}\right)\Bigg]+\dots \nonumber \\
\delta w_{1_{-1,0}}&=&\langle n'|\langle 1_{-1,0}|\delta\hat{H}|1_{-1,0}\rangle|n'\rangle,
\end{eqnarray}
where we assume $|n'\rangle$ to be diagonal in $\langle 1_{-1,0}|\delta\hat{H}|1_{-1,0} \rangle$. The components $\alpha_{XX}$, $\alpha_{YY}$, $\alpha_{ZZ}$, $B_{XX}$, $B_{YY}$ and $B_{ZZ}$ now make anisotropically weighted contributions reflecting the idea that the $X$ axis is most likely to be found perpendicular to the $x$-$y$ plane in the $1_{-1,0}$ rotor state: the electric and magnetic field vectors of the light can be said to drive oscillations less frequently along the $X$ axis and more frequently along the $Y$ and $Z$ axes. Such observations can be extended, of course, to other rotor and nuclear-spin states.  The important point here is that the energy shifts due to the light exhibit different dependencies upon $B_{XX}$, $B_{YY}$ and $B_{ZZ}$ for different rotor states whilst differing for opposite circular polarisations: the rotation and hence orientation of the molecule relative to the light differs for different rotor states whilst one enantiomorphic form of the helically twisting electric and magnetic field vectors that comprise circularly polarised light \cite{Takeda 14} is more competent at driving chiral oscillations in the charge and current distributions of the molecule than the other, much as one enantiomorphic form of a glove is a better fit for a human hand than the other. Similarly for a fixed circular polarisation and opposite enantiomers. In contrast the chirally sensitive phase that underpins chiral microwave three wave mixing derives from the sign of the product of three orthogonal electric-dipole moment components, which is opposite for opposite enantiomers \cite{Hirota 12, Nafie 13, Patterson 13, Patterson 13b, Shubert 14a, Shubert 14, Lobsiger 14, Lehman 15a, Shubert 15, Shubert 15 b}. The diagonalisation of $\hat{H}'$ is discussed in more detail in Appendix A.

\subsection{Observing and interpreting the response}
We envisage having a large number of molecules in practice, occupying many rotational and nuclear-spin states in accord with some thermal distribution, say. We recognise the need, therefore, to observe and interpret their chiroptical response in a manner that distinguishes between different rotational states, lest we lose the orientated character that is inherent to these states individually but absent from them collectively \cite{Vleck 32}. We propose simply measuring the rotational spectrum of the molecules in the microwave domain \cite{Cleeton 34}, which will appear modified due to the light. For example, the microwave energy required to induce a $1_{-1,0}\leftarrow 0_{0,0}$ rotational transition in a molecule follows from the difference between (\ref{G12b}) and (\ref{G12a}) as
\small
\begin{eqnarray}
B & + & C-\frac{I}{\epsilon_0 c} \Bigg[ -\frac{2}{15}\left(\frac{1}{2}\alpha_{XX}+\sigma |\mathbf{k}|B_{XX}\right) \nonumber \\
& + & \frac{1}{15}\left(\frac{1}{2}\alpha_{YY}+\sigma |\mathbf{k}|B_{YY}\right)  +\frac{1}{15}\left(\frac{1}{2}\alpha_{ZZ}+\sigma |\mathbf{k}|B_{ZZ}\right) \Bigg]+\dots \nonumber
\label{hello2}
\end{eqnarray}
\normalsize
plus a small correction moreover that is particular to the nuclear-spin states involved. $B_{XX}$, $B_{YY}$ and $B_{ZZ}$ can be determined individually by recording such energies for two distinct rotor transitions and both circular polarisations of the light and making use of the measured value of the isotropic sum (\ref{isotropic sum}). This is the essence of chiral rotational spectroscopy. Let us emphasise here, however, that chiral rotational spectroscopy also has abilities reaching beyond this particular task, as we will see in what follows.

\subsection{Additional remarks}
Knowledge of $B_{XX}$, $B_{YY}$ and $B_{ZZ}$ might assist in the assignment of absolute configuration, as the measured signs of these should be easier to correlate with those predicted by quantum chemical calculations than in the case of the isotropic sum (\ref{isotropic sum}), which is often somewhat smaller in magnitude than its constituents $B_{XX}/3$, $B_{YY}/3$ and $B_{ZZ}/3$ \cite{Zuber 08}. $B_{XX}$, $B_{YY}$ and $B_{ZZ}$ might also serve as probes of isotopic molecular chirality and cryptochirality in general, where the isotropic sum (\ref{isotropic sum}) fails rather dramatically, as we will elucidate in \S\ref{Isotopicsubsection}. Although our focus in the present paper is upon the chirality of individual molecules, we observe that knowledge of $B_{XX}$, $B_{YY}$ and $B_{ZZ}$ might in some cases facilitate the exploration and exploitation of the myriad contributions to the optical properties of crystals \cite{Kaminsky 00, Kahr 12} comprised, wholly or in part, of such molecules. We recognise moreover that our proposed technique offers $\alpha_{XX}$, $\alpha_{YY}$ and $\alpha_{ZZ}$ and potentially even the distortion of such quantities by static fields (see Appendix B) as by-products, which is in itself an attractive feature that could see our proposed technique find use even for achiral molecules.

It is interesting to note that $\hat{H}_\textrm{light}$ is, in fact, the a.c. Stark Hamiltonian, but calculated here to higher order than is usual \cite{Cameron 14 c}. The associated energy shifts are the same as those that govern the refraction of light propagating through a medium \cite{Cameron 14 c}, with circular birefringence due to $B_{XX}$, $B_{YY}$ and $B_{ZZ}$ giving rise to natural optical rotation \cite{Cameron 14 c}. Spatial gradients in such shifts give rise, moreover, to forces, including the dipole optical force used to trap atoms in optical lattices \cite{Metcalf 99} and the discriminatory optical force \cite{Cameron 14 c, Canaguier 13, Cameron 14 a, Wang 14, Cameron 14 b, Ding 14, Canaguier 14, Bradshaw 14, Chen 14, Alizadeh 15, Canaguier 15}: a viable manifestation of chirality in the translational degrees of freedom of chiral molecules. 

Let us conclude the present section now with a discussion of other phenomena and techniques centred upon the rotational degrees of freedom of chiral molecules, by way of comparison with chiral rotational spectroscopy. Microwave optical rotation and circular dichroism have been considered in theory  \cite{Salzman 77, Polavarapu 87, Salzman 87a, Salzman 87b, Salzman 89, Salzman 90a, Salzman 90b, Salzman 91a, Salzman 91b, Salzman 97, Salzman 98}. These phenomena promise chirally sensitive information about a molecule's permanent electric-dipole moment and rotational $g$ tensor but are anticipated to be weak, owing primarily to the smallness of molecules relative to the twist inherent to circularly polarised microwaves. Rotational Raman optical activity has also been considered in theory \cite{Polavarapu 87, Barron 85}. This phenomenon promises certain combinations of orientated polarisability components. A difficulty with rotational Raman optical activity is the anticipated proximity of the relevant Stokes and anti-Stokes lines to the Rayleigh line \cite{Barron 15}. In light of these challenges it is little surprise perhaps that ``\textit{no experimental observations ... of optical activity associated with pure rotational transitions of chiral molecules ... (had) been reported}" by 2004 \cite{Barron 04}. The successful implementation in 2013 of chiral microwave three wave mixing \cite{Hirota 12, Nafie 13, Patterson 13, Patterson 13b, Shubert 14a, Shubert 14, Lobsiger 14, Lehman 15a, Shubert 15, Shubert 15 b}, however, demonstrated that the exploitation of rotational degrees of freedom is, in fact, viable. Two additional works of interest came to our attention whilst preparing the present paper for submission. The first of these is a theoretical proposal for orientating chiral molecules using multi-coloured light \cite{Takemoto 08}. The second is a theoretical proposal, published on the arXiv, for the use of ``\textit{near-resonant AC Stark shifts}'' to detect molecular chirality in the microwave domain via a ``\textit{five wave mixing}'' process \cite{Lehman 15b}. Let us emphasise that chiral rotational spectroscopy is quite distinct from these techniques, including chiral microwave three wave mixing, and that it offers fundamentally different information about molecular chirality.


\section{Chiral rotational spectra}
\label{Chiralrotationalspectra}

In the present section our goal is to illustrate, simply, some of the features that might be seen in chiral rotational spectra for various different types of sample. To produce FIG. \ref{Alaninespectrumfigure},  FIG. \ref{Isotopicspectrumfigure}, FIG. \ref{Tartaricspectrumfigure} and FIG. \ref{Ibuprofenspectrumfigure} we plotted Lorentzians, centred at the relevant rotational transition frequencies as given by the leading-order perturbative results described in Appendix A but with $\delta\hat{H}$ neglected here. Each Lorentzian was ascribed a frequency full-width at half-maximum of $1.0\times10^3$s$^{-1}$ and taken to be proportional in amplitude to the number of contributing molecules. The same features persist when higher-order corrections and the effects of $\delta\hat{H}$ are included and for larger rotational linewidths: it is acceptable to have rotational lines overlap significantly if their centres, say, can still be distinguished with sufficient resolution. The forms of the rotational lines seen in a real chiral rotational spectrum will depend, of course, upon the nature and functionality of the chiral rotational spectrometer used to obtain the spectrum, but should nevertheless offer the same information. The calculated molecular properties used to produce FIG. \ref{Alaninespectrumfigure}, FIG. \ref{Isotopicspectrumfigure}, FIG. \ref{Tartaricspectrumfigure} and FIG. \ref{Ibuprofenspectrumfigure} are reported in Appendix C. The reader will observe the high precision with which $I$ and $2\pi/|\mathbf{k}|$ are quoted in the present section. In principle this represents no difficulty and ensures that FIG. \ref{Alaninespectrumfigure}, FIG. \ref{Isotopicspectrumfigure}, FIG. \ref{Tartaricspectrumfigure} and FIG. \ref{Ibuprofenspectrumfigure} are drawn accurately to a frequency resolution of $10^2$s$^{-1}$. In practice it should be possible in many cases to reduce stringent requirements on the uniformity and stability of the intensity of the light by exploiting certain, `magic' rotational transitions, as discussed in \S\ref{Practical}.

\subsection{Orientated chiroptical information}

\begin{figure}[h!]
\centering
\includegraphics[width=\linewidth]{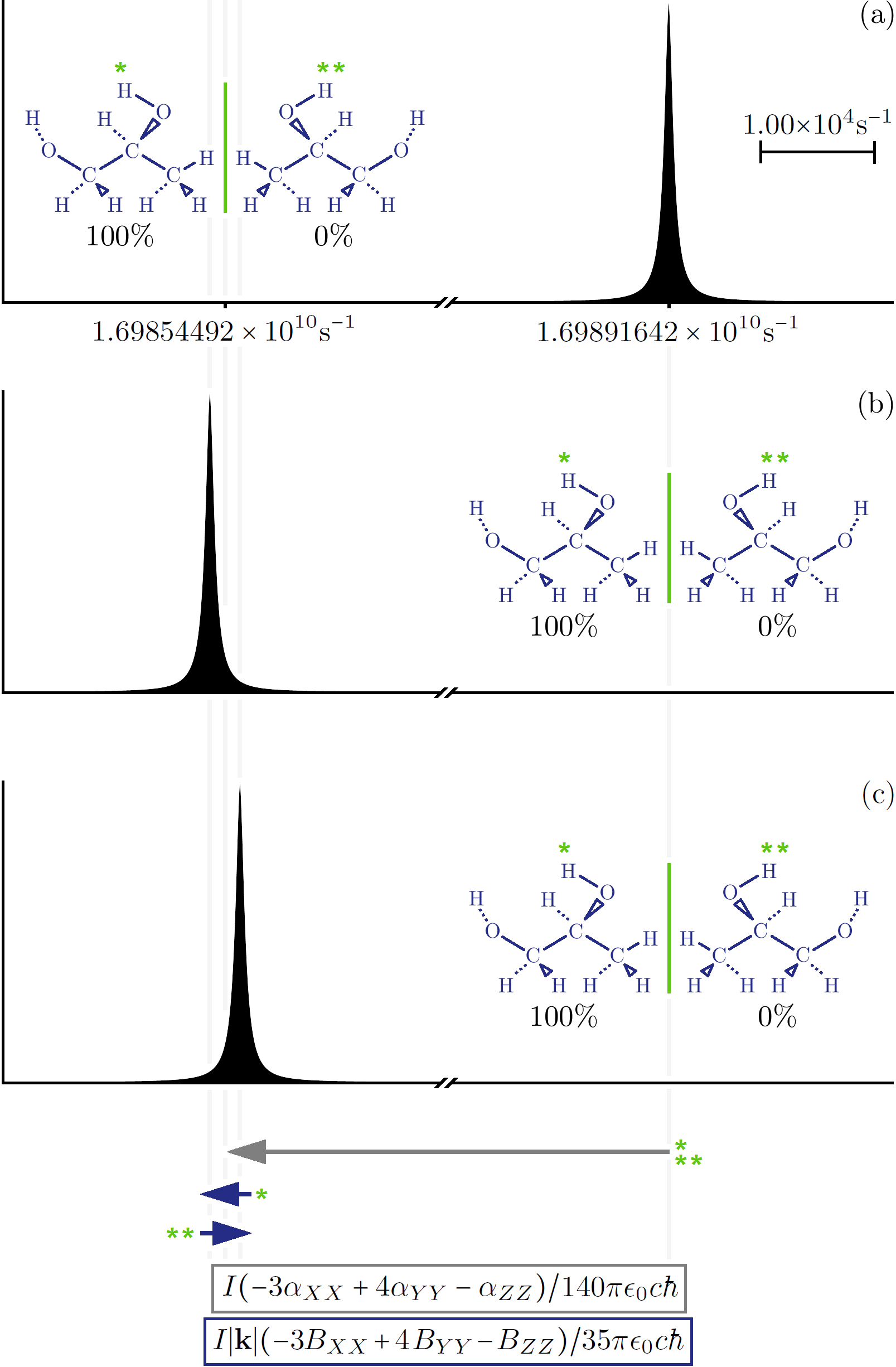}\\
\caption{\small A rotational line for an enantiopure sample of the lowest energy conformer of (S)-propylene glycol in the absence of light (a), illuminated by left-handed light (b) and illuminated by right-handed light (c). The separation between rotational lines (b) and (c) in particular reveals orientationally and chirally sensitive information about the response of the molecules to the light.}
\label{Alaninespectrumfigure}
\end{figure}

Consider first an enantiopure sample of the lowest energy conformer of (S)-propylene glycol \cite{Lovas 09}. Racemic propylene glycol is employed as an antifreeze and is a key ingredient in electronic cigarettes. Depicted in FIG. \ref{Alaninespectrumfigure} is: (a) the $2_{-1}\leftarrow 1_{-1}$ rotational line in the absence of light; (b) the $2_{-1,0}\leftarrow 1_{-1,0}$ rotational line in the presence of light with $I=2.0000\times10^{12}$kg.s$^{-3}$, $2\pi/|\mathbf{k}|=5.320\times10^{-7}$m and $\sigma=1$; (c) the same as in (b) but with $\sigma=-1$. The separation between rotational line (a) and the centroid of rotational lines (b) and (c) yields a certain combination of $\alpha_{XX}$, $\alpha_{YY}$ and $\alpha_{ZZ}$ whilst that between rotational lines (b) and (c) yields a certain combination of $B_{XX}$, $B_{YY}$ and $B_{ZZ}$, as described in \S\ref{Chiralrotationalspectroscopy}.

\subsection{Isotopic molecular chirality}
\label{Isotopicsubsection}

\begin{figure}[h!]
\centering
\includegraphics[width=0.6\linewidth]{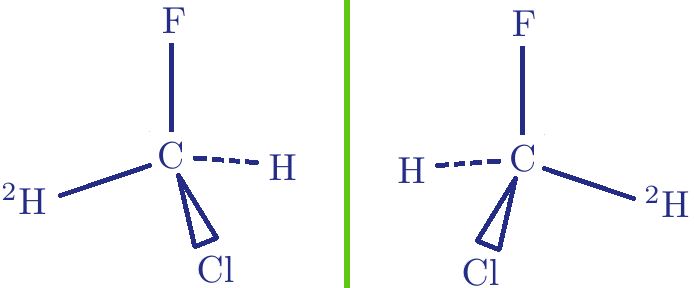}\\
\caption{\small Singly deuterated chlorofluoromethane derives chirality from the arrangement of its neutrons. These molecules should exist in small quantities in some refrigerators.} 
\label{Figure2}
\end{figure}

Chirality is more widespread at the molecular level than is sometimes appreciated, for even a molecule with an achiral arrangement of atoms may in fact be chiral solely by virtue of its isotopic constitution, as illustrated in FIG. \ref{Figure2}. Isotopically chiral molecules might have been amongst the very first chiral molecules, formed perhaps in primordial molecular clouds \cite{Oba 15}. They might even have given rise to biological homochirality, by triggering dissymmetric autocatalysis reactions \cite{Frank 53, Soai 95, Barabas 08, Kawasaki 09, Oba 15}. At a more fundamental level still, isotopically chiral molecules have been put forward \cite{Berger 05} as promising candidates for the measurement of minuscule differences believed to exist between the energies of opposite enantiomers \cite{Lee 56, Wu 57, Rein 74, Letokhov 75, Lough 02, Barron 04, Bunker 05}. It is well established that isotopic substitution in certain \textit{achiral} molecules can significantly modify their interaction with living things. Heavy water can change the phase and period of circadian oscillations \cite{Bruce 60}, for example. In spite of this there ``\textit{have been very few studies on isotope-generated chirality in biochemistry}'' \cite{Barabas 08}. 

\begin{figure}[h!]
\centering
\includegraphics[width=\linewidth]{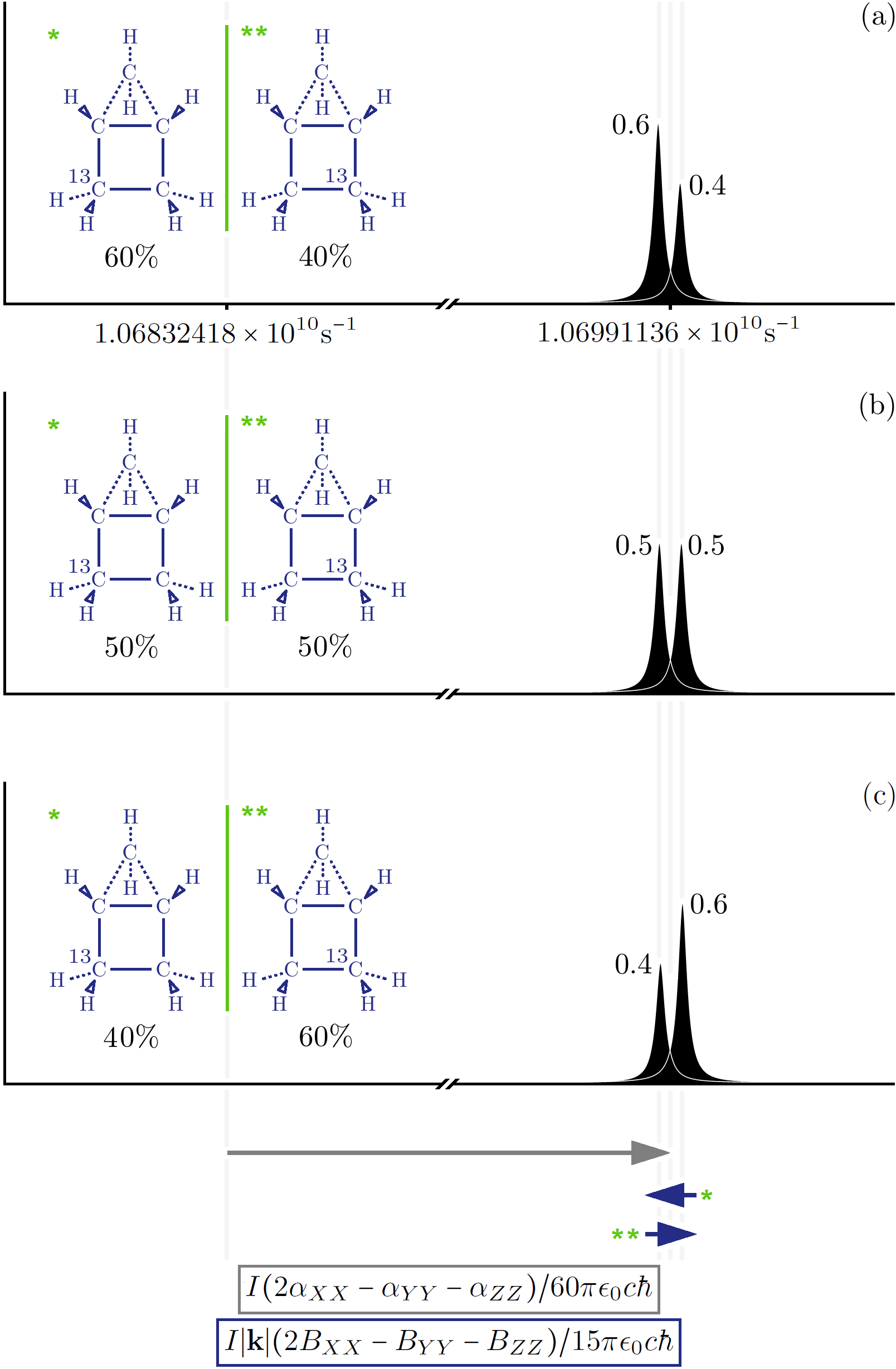}\\
\caption{\small A rotational line for a $60:40$ mixture (a), a $50:50$ mixture (b) and a $40:60$ mixture (c) of opposite enantiomers of isotopically chiral housane in the presence of left-handed light. The chiral splitting is apparent even for the racemate whilst the relative heights of the constituent lines reveal the enantiomeric excesses of the samples.} 
\label{Isotopicspectrumfigure}
\end{figure}

Isotopic molecular chirality can already be probed using various techniques, in particular vibrational circular dichroism and Raman optical activity, which are inherently sensitive to chiral mass distributions \cite{Holzwarth 74, Barron 77, Barron 78, Meddour 194}. A difficulty, however, is that enantiopurified samples of isotopically chiral molecules can often only be synthesised in small quantities \cite{Barron 78} whilst resolution of racemates is extremely challenging \cite{Kimata 97}. Chiral rotational spectroscopy may prove particularly useful here as it is, like vibrational circular dichroism and Raman optical activity, inherently sensitive to isotopic molecular chirality and, in addition, gives an incisive signal even for a racemate, thus \textit{negating} the need for dissymmetric synthesis or resolution

We find in electronic calculations within the Born-Oppenheimer approximation for a rigid nuclear skeleton \cite{Born 27, Bunker 05} that the isotropic sum (\ref{isotropic sum}) vanishes for an isotopically chiral molecule, as it is rotationally invariant and the electronic charge and current distributions of the molecule are achiral. Chirally sensitive vibrational corrections to this picture do exist but are usually small at visible or near infrared frequencies as considered here. The individual components $B_{XX}$, $B_{YY}$ and $B_{ZZ}$, and therefore chiral splittings in chiral rotational spectroscopy, can nevertheless attain appreciable magnitudes for an isotopically chiral molecule as each of these is dependent upon the orientation of the principal axes of inertia relative to the molecule and is, therefore, sensitive to the distribution of mass throughout the molecule, which is where the molecule's chirality resides. Chiral rotational spectroscopy might be similarly useful for other molecules exhibiting cryptochirality \cite{Mislow 77} where the isotropic sum (\ref{isotropic sum}) is essentially zero whilst two or three of its constituents $B_{XX}/3$, $B_{YY}/3$ and $B_{ZZ}/3$ are instead of appreciable magnitude. To the best of our knowledge the use of chiral microwave three wave mixing to probe molecules for which the chirality resides in an isotopic substitution has not yet been reported.

Consider next then a non-enantiopure sample of housane with the usual C atom at either the bottom-left or bottom-right of the `house' substituted with a $^{13}$C atom to give the opposite enantiomers of an isotopically chiral molecule. Depicted in FIG. \ref{Isotopicspectrumfigure} is the $1_{-1,0}\leftarrow 0_{0,0}$ rotational line for light with $I=4.0000\times10^{12}$kg.s$^{-3}$, $2\pi/|\mathbf{k}|=5.320\times10^{-7}$m and $\sigma=1$ illuminating a sample comprised of: (a) a $60:40$ mixture of opposite enantiomers; (b) a $50:50$ mixture; (c) a $40:60$ mixture. In all three cases the chiral splitting is apparent whilst the relative heights of the constituent lines reveal the enantiomeric excess of the sample and so enable its determination. Let us highlight the significance of panel (b) in particular. We have here an obvious and revealing signature of chirality from a racemate of isotopically chiral molecules, as claimed. The chirality of each of these molecules derives solely from the placement of a single neutron, which constitutes but $1\%$ of the total mass of the molecule. Techniques such as electronic optical rotation and electronic circular dichroism in contrast are nearly double blind under such circumstances and even vibrational circular dichroism, Raman optical activity and chiral microwave three wave mixing would yield vanishing signals.

\subsection{Molecules with multiple chiral centres}

\begin{figure}[h!]
\centering
\includegraphics[width=\linewidth]{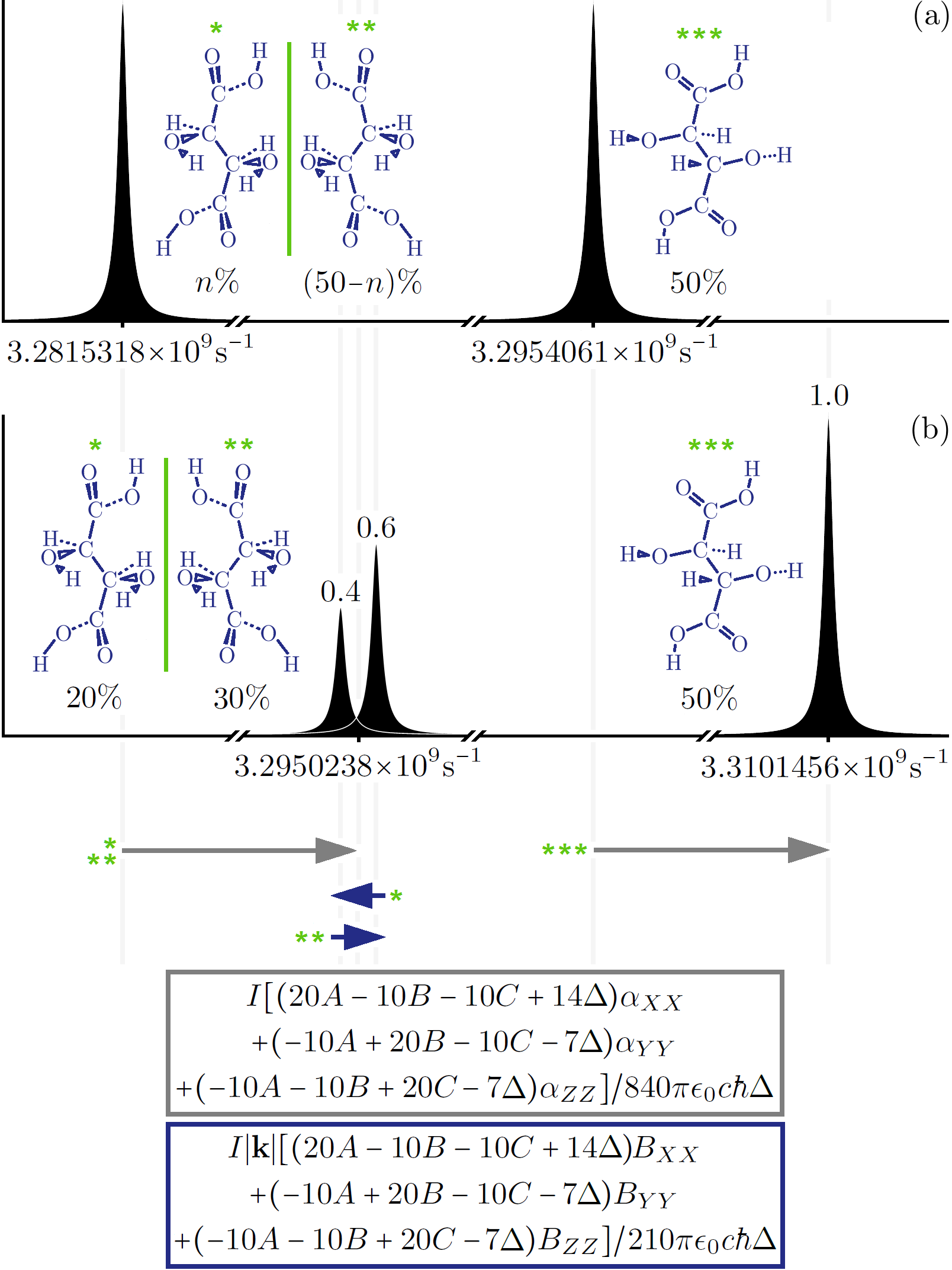}\\
\caption{\small A rotational line for a sample of the three different stereoisomers of tartaric acid as it might appear in standard rotational spectroscopy (a) and in chiral rotational spectroscopy using left-handed light (b). The standard rotational spectrum fails to distinguish between the opposite enantiomers whilst the chiral rotational spectrum instead distinguishes between all three molecular forms. Here $\Delta=\sqrt{A^2+B^2+C^2-AB-AC-BC}$.}
\label{Tartaricspectrumfigure}
\end{figure}

Standard rotational spectroscopy can often distinguish well between different isomers, provided they are \textit{not} opposite enantiomers \cite{Lovas 09}. Chiral rotational spectroscopy can distinguish well between different isomers \textit{including} opposite enantiomers. It may find particular use, therefore, in the analysis of molecules with multiple chiral centres, which permit an exponentially large number of different stereoisomers, many of which are opposite enantiomers. This in turn could see chiral rotational spectroscopy find particular use in the food and pharmaceutical industries, where different isomers must be individually justified \cite{EMICURE} and molecules with multiple chiral centres are recognised as being ``\textit{challenging}'' \cite{YAHOO}. 

Consider now then a sample of tartaric acid. The two chiral centres permit three different stereoisomers. One of these; mesotartaric acid, is achiral whilst the other two; $\textsc{L}$-tartaric acid and $\textsc{D}$-tartaric acid, are opposite enantiomers. $\textsc{L}$-tartaric acid is found in grapes and was one of the first molecules recognised as being optically active \cite{Barron 04}. The racemate of $\textsc{L}$- and $\textsc{D}$-tartaric acid, also known as paratartaric acid \cite{Barron 04} or racemic acid \cite{F7}, was the subject of Pasteur's original chiral separation \cite{Lough 02, Barron 04}. Depicted in FIG. \ref{Tartaricspectrumfigure} (a) is the $2_{-2}\leftarrow1_{-1}$ rotational line for a $50:n:(50-n)$ mixture of mesotartaric acid, $\textsc{L}$-tartaric acid and $\textsc{D}$-tartaric acid in the absence of light \cite{F8}. The contribution due to mesotartatic acid appears well separated from that due to $\textsc{L}$-tartaric acid and $\textsc{D}$-tartaric acid. The spectrum gives no information, however, about the relative abundances of $\textsc{L}$-tartaric acid and $\textsc{D}$-tartaric acid, only their combination. Depicted in FIG. \ref{Tartaricspectrumfigure} (b) is the $2_{-2,0}\leftarrow 1_{-1,\pm1}$ rotational line for a $50:20:30$ mixture in the presence of light with $I=1.0000\times10^{12}$kg.s$^{-3}$, $2\pi/|\mathbf{k}|=5.320\times10^{-7}$m and $\sigma=1$. Contributions due to all \textit{three} stereoisomers now appear well distinguished whilst yielding a wealth of new information, as claimed. 

Rotational spectra are sufficiently sparse that the analysis of molecules with significantly more chiral centres in this way should not be met with any fundamental difficulties. This ability to distinguish well and in a chirally sensitive manner between subtly different molecular forms persists moreover for more general mixtures containing multiple types of molecule. The chirally sensitive analysis of complicated mixtures using traditional techniques represents a serious challenge. Indeed, it was suggested in 2014 that ``\textit{only one mixture analysis (based upon circular dichroism, vibrational circular dichroism or Raman optical activity) was reported so far}'' \cite{Shubert 14a}, although the use of chiral microwave three wave mixing to analyse various mixtures has now been well demonstrated \cite{Shubert 14a, Shubert 14, Shubert 15, Shubert 15 b}.

\subsection{Scaling}

\begin{figure}[h!]
\centering
\includegraphics[width=\linewidth]{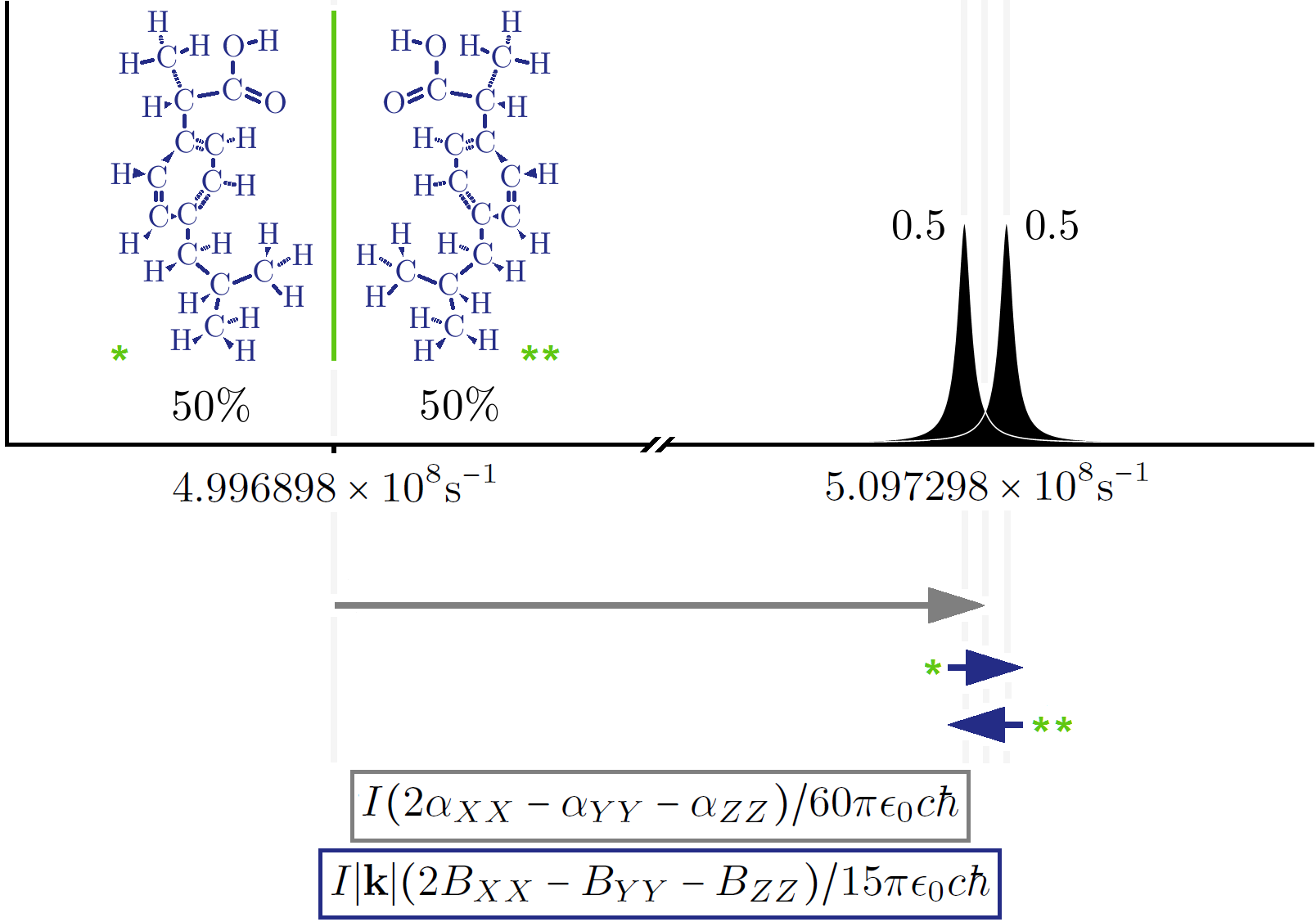}\\
\caption{\small A chiral splitting induced in a rotational line of a racemate of ibuprofen using a relatively low intensity of left-handed light.}
\label{Ibuprofenspectrumfigure}
\end{figure}

Polarisabilities tend to increase with the size of a molecule. The light intensity required to induce observable shifts in a rotational spectrum therefore tends to decrease with the size of a molecule, as is evident in the examples above. This favourable scaling is ultimately counteracted in that larger molecules are usually more difficult to sample appropriately, tend to exhibit lower rotational transition frequencies, are often more likely to absorb light and might require higher levels of theory to accurately describe. It seems then that there should be a certain molecular size range for which chiral rotational spectroscopy is particularly well suited. To illustrate these ideas let us consider a racemate of a particular conformer of ibuprofen, which is somewhat more massive than the molecules considered in the other examples above. Such a sample would yield no information about the chirality of the molecules when analysed using traditional techniques, in spite of the fact that it is only the (S)-enantiomeric form of ibuprofen that acts as the anti-inflammatory agent whilst the (R)-enantiomeric form is ineffective in this context \cite{Sarker 07}. Enantiopure ibuprofen is sometimes sold under a different name such as Seractil\circledR. Depicted in FIG. \ref{Ibuprofenspectrumfigure} is the chiral splitting of the $1_{-1,0}\leftarrow 0_{0,0}$ rotational line due to light with $I=2.0000\times10^{11}$kg.s$^{-3}$, $2\pi/|\mathbf{k}|=5.320\times10^{-7}$m and $\sigma=1$. The presence and chiral character of the opposite enantiomeric forms is revealed, with a light intensity considerably lower than in the other examples above, as claimed. The rotational transition frequencies seen here are also considerably lower than in the other examples above, although it should be noted that these are amongst the very lowest rotational transition frequencies available for these molecules and that significantly higher rotational transition frequencies do exist.

The chiral splittings seen in FIG. \ref{Alaninespectrumfigure},  FIG. \ref{Isotopicspectrumfigure}, FIG. \ref{Tartaricspectrumfigure} and FIG. \ref{Ibuprofenspectrumfigure} are neither the smallest nor the largest to be found in the chiral rotational spectra of these molecules.

\subsection{Practical considerations}
\label{Practical}
Requirements on the monochromaticity and stability of the wavelength of the light are stringent but are eased somewhat by the fact that the $\alpha_{AB}$ vary slowly with wavelength far off-resonance. For most rotational transitions requirements on the uniformity and stability of the intensity of the light are very stringent, as small variations in the intensity can easily overwhelm chiral splittings. In many cases rotational transitions can be found, however, for which the chirally insensitive piece of the rotational transition frequency shift due to the light is considerably smaller than is typical whilst the chirality sensitive piece remains appreciable. These magic rotational transitions should be particularly well suited to chiral rotational spectroscopy as they reduce requirements on the uniformity and stability of the intensity of the light. It should be possible moreover to significantly refine some magic transitions by fine-tuning the polarisation properties of the light or even the strength and direction of an applied static field. 


\section{Chiral rotational spectrometer}
\label{Chiralrotationalspectrometer}
In the present section we discuss a basic design for a chiral rotational spectrometer \cite{PATENT}. This represents but one of many conceivable possibilities for the implementation of chiral rotational spectroscopy: the ideas introduced in \S\ref{Chiralrotationalspectroscopy} and \S\ref{Chiralrotationalspectra} have a generality reaching beyond the present discussions. The design certainly has its limitations, but should nevertheless permit high-precision measurements based upon $\alpha_{XX}$, $\alpha_{YY}$ and $\alpha_{ZZ}$ for many types of molecule, be they chiral or achiral, as well as measurements based upon $B_{XX}$, $B_{YY}$ and $B_{ZZ}$ for some types of chiral molecule under favourable circumstances.

\begin{figure}[h!]
\centering
\includegraphics[width=\linewidth]{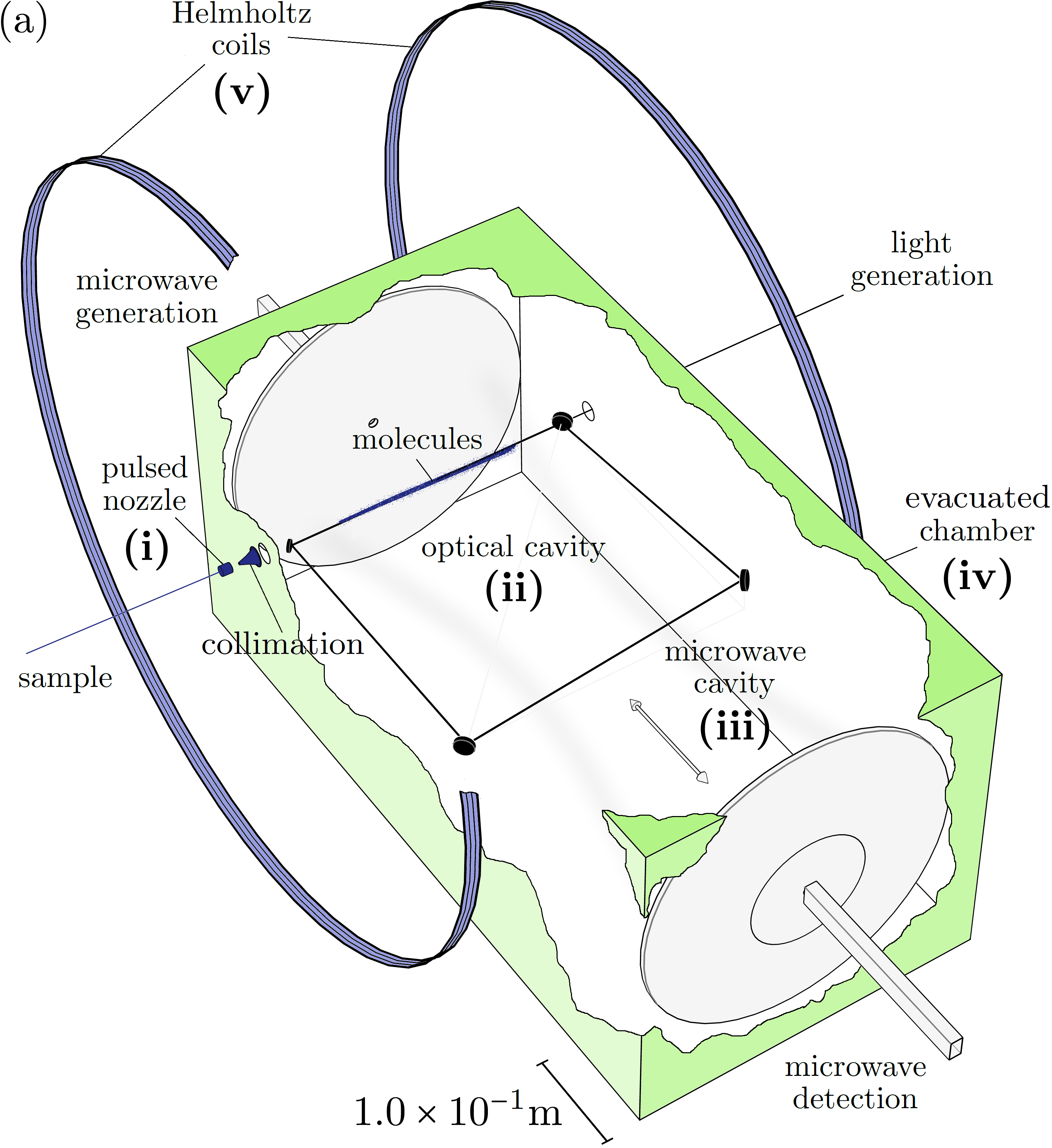} \\ 
\vspace{1cm}
\includegraphics[width=\linewidth]{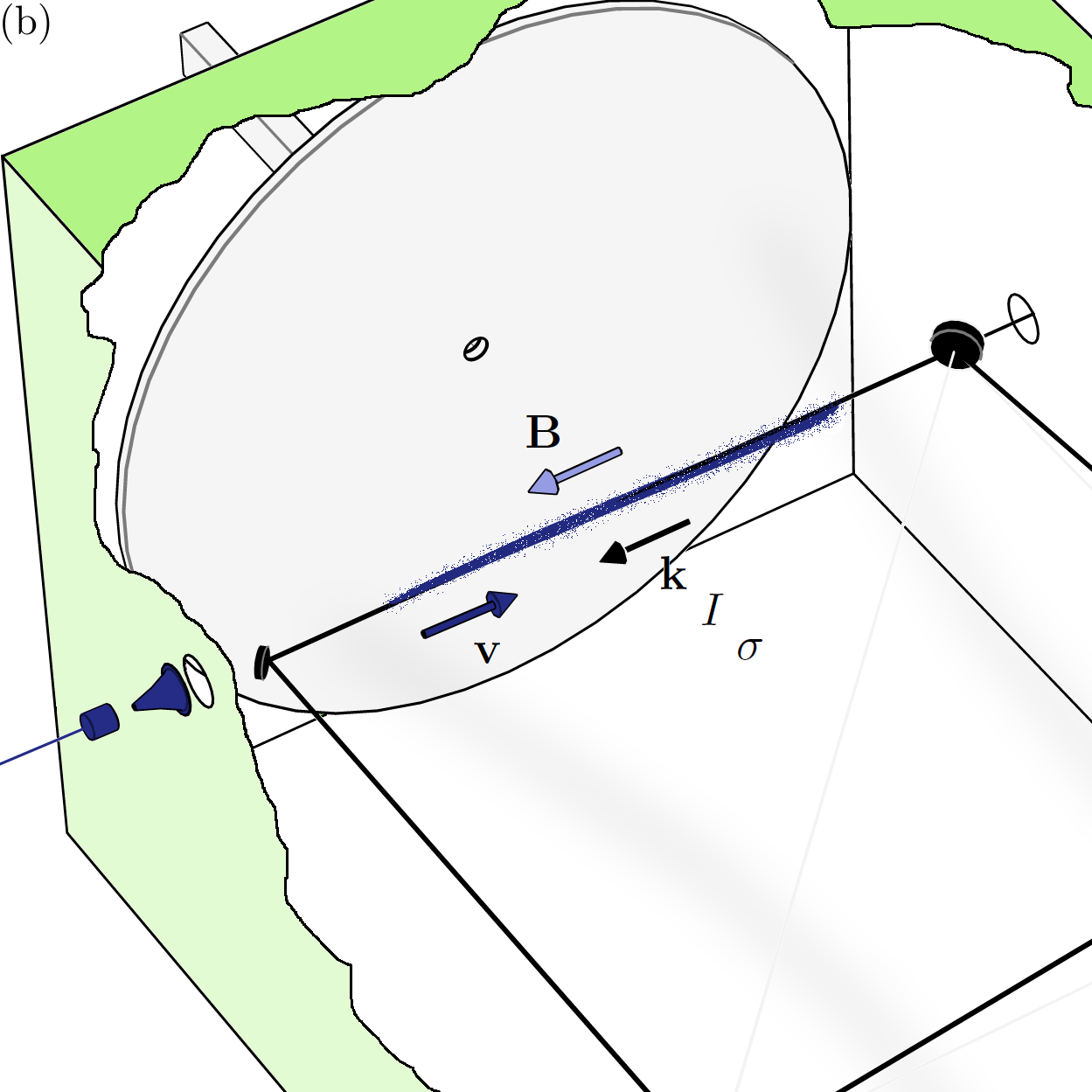} \\
\caption{\small Key components of a chiral rotational spectrometer. Drawn approximately to scale but with portions of the Helmholtz coils and the vacuum chamber removed, for the sake of clarity.} 
\label{CRSprot}
\end{figure}

The key components of the spectrometer are depicted in FIG. \ref{CRSprot} (a), with an expanded view of the active region in FIG. \ref{CRSprot} (b). We summarise their functionality as follows. A quantitative model of the spectrometer is given in Appendix D.
\begin{enumerate}

\item[\textbf{(i)}] A pulsed supersonic expansion nozzle together with a collimation stage is employed to generate narrow pulses of internally cold chiral molecules with unimpeded rotational degrees of freedom. A nozzle of the piezoelectric variety permits a high rate of measurement \cite{Cross 82}. The collimation stage might include a skimmer augmented by an aperture \cite{Kantrowitz 51, Kistiakowsky 51, Morse 96}. 

\item[\textbf{(ii)}] An optical cavity houses far off-resonance visible or perhaps near infrared circularly polarised light of moderate intensity, to shift the rotational energies of the molecules in a chirally sensitive manner. Fine-tuning the polarisation properties of the light in the active region enables the refinement of magic transitions, to help overcome stringent requirements on the intensity of the light and thus obtain a clean chiral rotational spectrum. The optical cavity might be of the skew-square ring variety, comprised of low-loss, ultra-high-reflectivity, low-anisotropy mirrors \cite{RingBook, Bilger 90} whilst the light might originate from an external cavity diode laser, the output of which is fibre amplified and mode matched into the ring with stability actively enforced \cite{Meng 05, Gold 14}. We envisage the light to be continuous wave here, with a central intensity of at least $10^{11}$kg.s$^{-3}$ ($10^7$W.cm$^{-2}$). Each molecule takes some $10^{-3}$s to traverse the light; a time interval large enough to facilitate a notional microwave frequency linewidth of around $10^4$s$^{-1}$. Variants of our design that use pulsed light rather than continuous wave light are also conceivable and may prove easier to implement in practice. We will discuss these in more detail elsewhere.

\item[\textbf{(iii)}] A microwave cavity and associated components generate and detect microwaves as in the well-established technique of cavity enhanced Fourier transform microwave spectroscopy \cite{Ekkers 76, Balle 79, Balle 80, Balle 81, Campbell 81 a, Campbell 81 b, Legon 83, Harmony 95, Suenram 99, Brown 03, Lovas 09} but here with the aim of measuring chirally sensitive distortions of the rotational spectrum of the molecules due to the light. The microwave cavity might be of the Fabry-P\'{e}rot variety, comprised of spherical mirrors with microwaves coupled in and out of the microwave cavity via waveguide or perhaps via antennas \cite{Balle 79, Balle 80, Balle 81, Campbell 81 a, Campbell 81 b, Legon 83, Harmony 95, Suenram 99, Brown 03, Lovas 09}. 

\item[\textbf{(iv)}] An evacuated chamber encompasses the key components described above to eliminate atmospheric interference with the molecular pulses and facilitate the removal of molecules between measurements. The absence of air, dust and other such influences should assist moreover in maintaining the stability of the optical cavity \cite{Gold 14}. 

\item[\textbf{(v)}] A static magnetic field of moderate strength and high uniformity defines a quantisation axis parallel to that defined by the direction of propagation of the light in the active region whilst enabling additional refinement of magic transitions if necessary. The static magnetic field might be produced by a pair of superconducting Helmholtz coils \cite{Read 82} or perhaps even an appropriate arrangement of permanent magnets with some degree of tunability. Note that the static magnetic field plays no direct role in probing the chirality of the molecules. Its influence is discussed in more detail in Appendix B.

\end{enumerate}
A chiral rotational spectrum is recorded as the average of many measurements, each of which proceeds as follows. The nozzle is opened at some initial time, allowing a molecular pulse to begin expanding towards the active region. In the initial stage of this expansion the internal temperature of the molecules decreases dramatically, as collisions convert enthalpy into directed translational energy. The molecules thus occupy their electronic and vibrational ground states and a small collection of rotational and nuclear-spin states, with their internal angular momenta preferentially quantised parallel to the static magnetic field. Following this initial stage the molecules proceed largely collision free. A subset of the molecules selected by the collimation stage eventually permeate the light in the active region, which shifts their rotational energies in a chirally sensitive manner. When the overlap between the molecules, the light and the microwave mode is optimum a microwave pulse permeates the microwave cavity and induces coherence in those (light-shifted) rotational transitions that lie near the chosen microwave cavity frequency and within the microwave cavity frequency bandwidth. The molecules then radiate back into the microwave cavity over a longer time, with the signal diminishing primarily as a result of residual collisions. This free induction decay signal is monitored and the real part of its Fourier transform, say, calculated and regarded as the measurement.

In Appendix \ref{SNRappendix} we estimate the signal-to-noise ratio and find that a very agreeable chiral rotational spectrum could be obtained for a recording time of a few hours under favourable operating conditions. This is approaching the time usually taken to record a complete standard rotational spectrum \cite{Lovas 09}, but here with the effort focused entirely upon a single rotational line. This is acceptable as four spectra spread over two lines for opposite circular polarisations might already permit the extraction of all of the chirally sensitive information on offer here for a particular enantiomer. We are reminded of early Raman optical activity spectrometers, which demanded recording times of several hours \cite{Barron 07}. Even now, ``\textit{traditional chiroptical spectroscopy techniques take minutes to hours}'' \cite{Nafie 13}. Chiral microwave three wave mixing in contrast exhibits an excellent signal-to-noise ratio, with measurement times ``\textit{as fast as tens of seconds}'' having been claimed in one of the earliest publications \cite{Nafie 13}.

A linearly polarised standing wave of light with a significantly lower intensity, housed simply in a two-mirror optical cavity perhaps, might already suffice if measurements based upon $\alpha_{XX}$, $\alpha_{YY}$ and $\alpha_{ZZ}$, for either chiral or achiral molecules, are all that is sought. \textit{Note added post-publication}: measurements of individual, orientated components of $\alpha_{AB}$ using a combination of optically induced ac Stark shifts and microwaves have recently been reported for heteronuclear molecules \cite{Gregory17a}. This work, performed independently of ours, confirms the validity of the basic theory first presented by us for chiral rotational spectroscopy.


\section{Summary and outlook} 
\label{Outlook}
In the present paper we have introduced chiral rotational spectroscopy: a new technique for chiral molecules that combines the chiral sensitivity inherent to natural optical activity with the orientational sensitivity and high precision inherent to standard rotational spectroscopy. Chiral rotational spectroscopy enables the determination of the orientated optical activity pseudotensor components $B_{XX}$, $B_{YY}$ and $B_{ZZ}$ of chiral molecules, in a manner that reveals the enantiomeric constitution of a sample and provides an incisive signal even for a racemate. It could find use in the analysis of molecules that are chiral solely by virtue of their isotopic constitution, molecules with multiple chiral centres and more besides.

There is much to be done: our formalism and calculations can and should be refined; the nature of the information offered by $B_{XX}$, $B_{YY}$ and $B_{ZZ}$ requires further attention; the use of our proposed technique to determine other polarisability components remains to be explored in more detail; designs for chiral rotational spectrometers and their functionality demand further investigation. 
We will return to these and related tasks elsewhere.


\section{Acknowledgements} 
\label{Ack}
This work was supported by the Engineering and Physical Sciences Research Council grants EP/M004694/1, EP/101245/1 and EP/M01326X/1; the alumnus programme of the Newton International Fellowship and the Max Planck Institute for the Physics of Complex Systems. We thank Melanie Schnell, Laurence D. Barron and Fiona C. Speirits for helpful correspondences.



\begin{appendix}
\section{Diagonalisation of $\hat{H}'$}
\label{DiagonalAppendix}
In the present appendix we discuss the diagonalisation of $\hat{H}'$ in more detail. We again focus our attention upon a molecule with nuclear spins of $0$ or $1/2$, assume that $\hat{H}_\textrm{rotor}\gg\hat{H}_\textrm{light}\gg\delta \hat{H}$ with no accidental degeneracies of importance and neglect the possibility of effects due to the spin statistics of similar nuclei \cite{Townes 55, Ramsey 56, Bunkers 98, Atkins 11}. For a molecule with nuclear spins of $1$ or greater the interaction between nuclear electric-quadrupole moments and intramolecular electric field gradients \cite{Bragg 48, Bragg 49, Townes 55, Ramsey 56} might give rise to a large $\delta\hat{H}$ such that $\hat{H}_\textrm{rotor}\gg\hat{H}_\textrm{light}\gg\delta \hat{H}$ is not a valid assumption and a more involved approach towards diagonalisation than that described here is required. 

We begin by considering $\hat{H}_\textrm{rotor}$ in isolation. Let us introduce here the familiar symmetric rotor states $|J,K,m\rangle$, with $K\in\{0,\dots,\pm J\}$ determining the $Z$ component of the (oblate) rotor's angular momentum, say \cite{Townes 55, Bunker 05, Bernath 05, Atkins 11}. We expand the $|J_{\tau,m}\rangle$ in terms of these as
\begin{equation}
| J_{\tau,m}\rangle=\sum_{K=-J}^J \tilde{a}_{J,\tau}(K) |J,K,m\rangle.
\end{equation}
Closed forms for the $\tilde{a}_{J,\tau}(K)$ are not known at present. It has been established \cite{Wang 29, Bunker 05}, however, that the $(2J+1)\times(2J+1)$ matrix of $\hat{H}_\textrm{rotor}$ for given values of $J$ and $m$ can be partitioned into smaller blocks, referred to as the $E^+$, $E^-$, $O^+$ and $O^-$ blocks with associated basis states
\begin{eqnarray}
E^+:&& \  |J,0,m\rangle  \ \textrm{together with} \nonumber \\
&&\ \frac{1}{\sqrt{2}}(|J,K,m\rangle +|J,-K,m\rangle) \  (K \ \textrm{even},\ \ge 2), \nonumber \\
E^-:&&  \  \frac{1}{\sqrt{2}}(|J,K,m\rangle-|J,-K,m\rangle) \ (K \ \textrm{even},\  \ge 2), \nonumber \\
O^+:&&  \ \frac{1}{\sqrt{2}}(|J,K,m\rangle+|J,-K,m\rangle) \  (K \ \textrm{odd},\ \ge 1) \nonumber \\
O^-:&&  \ \frac{1}{\sqrt{2}}(|J,K,m\rangle-|J,-K,m\rangle) \ (K \ \textrm{odd},\  \ge 1). \nonumber
\end{eqnarray}
The $\tilde{a}_{J,\tau}(K)$ can then be found by diagonalising these blocks individually, the associated eigenvalues being the $w_{J_{\tau}}$ with $\tau$ running from $-J$ to $J$ with increasing energy. For the lowest values of $J$ this procedure can be performed analytically. For higher values of $J$ the $E^+$, $E^-$, $O^+$ and $O^-$ blocks must themselves be diagonalised numerically. In what follows we focus our attention upon a particular pair of values of $J$ and $\tau$. We assume the associated $\tilde{a}_{J,\tau}(K)$ to be known and that these satisfy $\sum_{K=-J}^J |\tilde{a}_{J,\tau}(K)|^2=1$, thus ensuring normalisation of the $|J_{\tau,m}\rangle$.

Next, we consider the perturbation of $\hat{H}_\textrm{rotor}$ by $\hat{H}_\textrm{light}$ to first order. The (2$J$+1)-fold $m$ rotational degeneracy inherent to $\hat{H}_\textrm{rotor}$ is partially broken by $\hat{H}_\textrm{light}$, as
\begin{eqnarray}
&&\langle J_{\tau,m}| \hat{H}_\textrm{light}|J_{\tau,m'}\rangle =  \\
&& - \delta_{mm'}   \frac{I}{\epsilon_0 c}\Bigg[ \mathtt{a}_{J,\tau}(|m|)\left(\frac{1}{2}\alpha_{XX}+\sigma |\mathbf{k}|B_{XX}\right) \nonumber \\
&&+\mathtt{b}_{J,\tau}(|m|)\left(\frac{1}{2}\alpha_{YY}+\sigma |\mathbf{k}|B_{YY}\right) \nonumber \\
&& +\mathtt{c}_{J,\tau}(|m|) \left(\frac{1}{2}\alpha_{ZZ}+\sigma |\mathbf{k}|B_{ZZ}\right) \Bigg] \nonumber
\end{eqnarray}
with
\begin{eqnarray}
\mathtt{a}_{J,\tau}(|m|)&=& \frac{1}{2}\sum_{K=-J}^J\sum_{K'=-J}^J \tilde{a}^\ast_{J,\tau}(K)\tilde{a}_{J,\tau}(K') \label{MF1} \\
&&\langle J,K,|m|| \left(\hat{\ell}_{xX}^2+\hat{\ell}_{yX}^2\right)|J,K',|m|\rangle, \nonumber \\
\mathtt{b}_{J,\tau}(|m|)&=& \frac{1}{2}\sum_{K=-J}^J\sum_{K'=-J}^J \tilde{a}^\ast_{J,\tau}(K)\tilde{a}_{J,\tau}(K') \label{MF2} \\
&& \langle J,K,|m|| \left(\hat{\ell}_{xY}^2+\hat{\ell}_{yY}^2\right)|J,K',|m|\rangle \nonumber \\
\mathtt{c}_{J,\tau}(|m|)&=& \frac{1}{2}\sum_{K=-J}^J\sum_{K'=-J}^J \tilde{a}^\ast_{J,\tau}(K)\tilde{a}_{J,\tau}(K') \label{MF3} \\
&&\langle J,K,|m|| \left(\hat{\ell}_{xZ}^2+\hat{\ell}_{yZ}^2\right)|J,K',|m|\rangle \nonumber
\end{eqnarray}
being numbers that quantify the average orientation of the molecule. The independence upon the sign of $m$ indicated here leaves a $|m|$ rotational degeneracy, of course, and may be appreciated by noting that a parity inversion of the system changes the sign of the component of the molecule's angular momentum along the direction of propagation of the light whilst leaving the energy of the system unchanged. The absence at this order of certain components such as $\alpha_{XY}$ may be appreciated by noting that these are not uniquely defined in the present context: a rotation of the molecular axes by $\pi$ about the original $X$ axis without changing the molecule leaves $\hat{H}_\textrm{rotor}$ unaffected whilst nevertheless changing the sign of $\alpha_{XY}$, for example. It is tedious but straightforward to evaluate the matrix elements appearing in (\ref{MF1}), (\ref{MF2}) and (\ref{MF3}), by performing angular integrations over direction cosines and symmetric rotor wavefunctions explicitly \cite{Bunker 05, Bernath 05} or by multiplying well-established expressions for direction cosine matrix elements in the symmetric rotor basis perhaps \cite{Eshbach 52, Townes 55}. We refrain from reproducing here the somewhat lengthy expressions thus obtained. We note, however, that the summations
\begin{eqnarray}
\mathtt{a}_{J,\tau}(|m|)+\mathtt{b}_{J,\tau}(|m|)+\mathtt{c}_{J,\tau}(|m|)&=&1, \\
\frac{1}{(2J+1)}\sum_{m=-J}^J \mathtt{a}_{J,\tau}(|m|)&=& \frac{1}{3}, \\
\frac{1}{(2J+1)}\sum_{m=-J}^J\mathtt{b}_{J,\tau}(|m|)&=& \frac{1}{3} \\
\frac{1}{(2J+1)}\sum_{m=-J}^J \mathtt{c}_{J,\tau}(|m|)&=& \frac{1}{3} 
\end{eqnarray}
yield isotropic values as indicated, in accord with the principle of spectroscopic stablity \cite{Vleck 32}. Higher-order corrections in the $\alpha_{AB}$ can be significant, but are chirally insensitive. We refrain, therefore, from including them explicitly in the present paper. Their presence is indicated in \S\ref{Chiralrotationalspectroscopy} by dots and is neglected in \S\ref{Chiralrotationalspectra}.

Finally, we consider the additional perturbation of $\hat{H}_\textrm{rotor}+\hat{H}_\textrm{light}$ by $\delta\hat{H}$ to first order. Let us introduce here the nuclear-spin states $|I_j,m_j\rangle$, with $I_j\in\{0,1/2\}$ determining the magnitude of the spin and $m_j\in\{0,\dots,I_j\}$ determining the $z$ component of the spin for the $j$th nucleus \cite{Townes 55, Ramsey 56, Bunker 05, Atkins 11}. Our approach is to consider each distinct value of $|m|=|m'|\in\{0,\dots,J\}$ in turn and diagonalise the matrix with elements of the form
\begin{equation}
\left(\prod_j  \langle I_j, m_j |\right)\langle J_{\tau,m} | \delta \hat{H} | J_{\tau,m'} \rangle\left(\prod_{j'}  |I_{j'},m'_{j'}\rangle\right). \nonumber
\end{equation}
The energy shifts thus obtained give rise in particular to hyperfine structure in the rotational spectrum of the molecule in the presence of the light. 

The leading-order perturbative results described above suffice to illustrate the basic features of chiral rotational spectroscopy and are the ones upon which we base our explicit discussions and calculations in the present paper. We note here, however, that near degeneracies of importance are, in fact, rather common. In general then, $\hat{H}'$ should be diagonalised numerically.


\section{Influence of an applied static magnetic field}
In the present appendix we briefly discuss the influence of an applied static magnetic field. We consider the situation described by $\hat{H}'$ as in Appendix A but augmented here by a uniform, static magnetic field $\mathbf{B}$ of moderate strength pointing in the $z$ direction. The rotational and nuclear-spin degrees of freedom of the molecule should now be well described by the effective Hamiltonian
\begin{eqnarray}
\hat{H}''&=&\hat{H}_\textrm{rotor}+\hat{H}_\textrm{light}+\hat{H}_\textrm{nucl}^\mathbf{B}+\hat{H}_\textrm{rotor}^\mathbf{B} \label{Franceisgreat} \\
&+&\hat{H}^\mathbf{B^2}+\hat{H}_\textrm{light}^\mathbf{B}+\delta\hat{H}+\delta\hat{H}^\mathbf{B}  \nonumber
\end{eqnarray}
with
\begin{equation}
\hat{H}_\textrm{nucl}^\mathbf{B}=- \frac{\mu_N}{\hbar} \sum_j g_j \hat{I}_z^j B_z
\end{equation} 
accounting for the interaction energy between $\mathbf{B}$ and the nuclear magnetic-dipole moments \cite{Townes 55, Ramsey 56},
\begin{equation}
\hat{H}_\textrm{rotor}^\mathbf{B}=-\frac{\mu_N}{2\hbar}\left(\hat{\ell}_{zA} \hat{J}_B +\hat{J}_B\hat{\ell}_{zA}\right)g_{AB} B_z 
\end{equation} 
accounting for the interaction energy between $\mathbf{B}$ and the rotational magnetic-dipole moment \cite{Eshbach 52, Burke 53, Flygare 74},
\begin{eqnarray}
\hat{H}^{\mathbf{B}^2}=-\frac{1}{2}\hat{\ell}_{zA}\hat{\ell}_{zB}\chi_{AB}B_z^2
\end{eqnarray}
accounting for the interaction energy between $\mathbf{B}$ and the magnetic-dipole moment induced by $\mathbf{B}$,
\begin{eqnarray}
\hat{H}_\textrm{light}^\mathbf{B}=-\frac{I }{2\epsilon_0 c} \frac{\sigma k_z  }{|\mathbf{k}|}   \hat{\ell}_{xA}\hat{\ell}_{yB}\hat{\ell}_{zC} \alpha'_{AB,C} B_z
\end{eqnarray} 
accounting for the distortion by $\mathbf{B}$ of the electronic electric-dipole / electric-dipole polarisability \cite{Cameron 14 c} and $\delta\hat{H}^\mathbf{B}$ accounting for additional effects associated with $\mathbf{B}$ such as nuclear-spin shielding. 
$\mu_N$ is the nuclear magneton; $g_j$ is the $g$ factor of the $j$th nucleus; $\hat{I}_z^j$ is the $z$ component of the spin of the $j$th nucleus; $B_z$ is the $z$ component of $\mathbf{B}$; the $g_{AB}$ are components of the rotational $g$ tensor, which has nuclear and electronic contributions \cite{Eshbach 52, Burke 53, Flygare 74}; the $\chi_{AB}$ are components of the electronic static magnetic susceptibility tensor, which has diamagnetic and temperature-independent paramagnetic contributions \cite{Vleck 32, Flygare 74, Barron 04}, and the $\alpha'_{AB,C}$ are components of the electronic Faraday-B polarisability \cite{Barron 04}. 

The $\alpha_{AB,C}'$ might in some cases give a $\sigma$-dependent contribution to $\hat{H}''$ comparable to that from the $B_{AB}$. Note, however, that the $\alpha'_{AB,C}$ are chirally insensitive: $\alpha'_{YZ,X}$, $\alpha_{ZX,Y}'$ and $\alpha'_{XY,Z}$ in particular are identical for opposite enantiomers. In principle the effects of the $\alpha'_{AB,C}$ can be distinguished from those of the $B_{AB}$ by comparing spectra obtained with $\mathbf{k}$ and $\mathbf{B}$ parallel and antiparallel. Indeed, the contribution made to $\hat{H}''$ by the $\alpha'_{AB,C}$ is to magnetic or Faraday optical rotation and the spin of light what the contribution made by the $B_{AB}$ is to natural optical rotation and the helicity of light \cite{Cameron 14 c}. 

We begin by considering the perturbation of $\hat{H}_\textrm{rotor}$ by $\hat{H}_\textrm{light}+\hat{H}_\textrm{nucl}^\mathbf{B}+\hat{H}_\textrm{rotor}^\mathbf{B}+\hat{H}^{\mathbf{B}^2}+\hat{H}_\textrm{light}^\mathbf{B}$ to first order. The nuclear-spin degeneracy is at least partially broken by $\hat{H}^\mathbf{B}_\textrm{nucl}$, as 
\begin{eqnarray}
\lefteqn{\left(\prod_j  \langle I_j, m_j |\right)\hat{H}_\textrm{nucl}^\mathbf{B}\left(\prod_{j'} |I_{j'},m'_{j'}\rangle\right)}\\
& = & -\left(  \prod_{j} \delta_{m_jm'_j}  \right)    \mu_N\left(\sum_{j'} g_{j'} m_{j'}\right)B_z  \nonumber 
\end{eqnarray}
with nuclear-spin degeneracies remaining when multiple nuclei of the same type with spins of $1/2$ are present, which will usually be the case. The $(2J+1)$-fold $m$ rotational degeneracy inherent to $\hat{H}_\textrm{rotor}$ is fully broken by $\hat{H}_\textrm{rotor}^\mathbf{B}$, as
\begin{equation}
\langle J_{\tau,m} |   \hat{H}_\textrm{rotor}^{\mathbf{B}}  | J_{\tau,m'}\rangle=-\delta_{mm'}\mu_N g_{J,\tau}m B_z
\end{equation}
with
\begin{eqnarray}
g_{J,\tau} m&=&\frac{1}{2\hbar} \langle J_{\tau,m}| \left(\hat{\ell}_{zX}\hat{J}_X+\hat{J}_X\hat{\ell}_{zX}\right)| J_{\tau,m} \rangle g_{XX}  \\
&&+\frac{1}{2\hbar} \langle J_{\tau,m}| \left(\hat{\ell}_{zY}\hat{J}_Y+\hat{J}_Y\hat{\ell}_{zY}\right)| J_{\tau,m} \rangle g_{YY} \nonumber \\
&&+\frac{1}{2\hbar} \langle J_{\tau,m}| \left(\hat{\ell}_{zZ}\hat{J}_Z+\hat{J}_Z\hat{\ell}_{zZ}\right)| J_{\tau,m} \rangle g_{ZZ} \nonumber
\end{eqnarray}
defining the effective rotational $g$ factor $g_{J,\tau}$ \cite{Eshbach 52, Burke 53}. Further $|m|$-dependent energy shifts arise through $\hat{H}^{\mathbf{B}^2}$, as
\begin{eqnarray}
\lefteqn{\langle J_{\tau,m}|\hat{H}^{\mathbf{B}^2}| J_{\tau,m'} \rangle} \nonumber \\*
& = & -\delta_{mm'} \frac{1}{2}\Big\{\left[1-2\mathtt{a}_{J,\tau}(|m|)\right] \chi_{XX} \\*
& \quad & + \left[1-2\mathtt{b}_{J,\tau}(|m|)\right] \chi_{YY}+\left[1-2\mathtt{c}_{J,\tau}(|m|)\right] \chi_{ZZ} \Big\}B_z^2, \nonumber 
\end{eqnarray}
and through $\hat{H}_\textrm{light}^\mathbf{B}$, as
\begin{eqnarray}
\lefteqn{\langle J_{\tau,m} |  \hat{H}_\textrm{light}^\mathbf{B} | J_{\tau,m'} \rangle} \nonumber \\*
& = & -\delta_{mm'}\frac{I}{2\epsilon_0 c}\frac{\sigma k_z }{|\mathbf{k}|}\Big\{ \left[2\mathtt{b}_{J,\tau}(|m|)+2\mathtt{c}_{J,\tau}(|m|)-1\right]\alpha_{YZ,X}' \nonumber  \\*
& \quad & + \left[2\mathtt{c}_{J,\tau}(|m|)+2\mathtt{a}_{J,\tau}(|m|)-1\right]\alpha_{ZX,Y}' \\*
& \quad & + \left[2\mathtt{a}_{J,\tau}(|m|)+2\mathtt{b}_{J,\tau}(|m|)-1\right]\alpha_{XY,Z}' \Big\}B_z, \nonumber 
\end{eqnarray}
\normalsize
although the magnitudes of these are not necessarily larger than those of the energy shifts that arise through $\delta\hat{H}+\delta\hat{H}^\mathbf{B}$. 

We conclude by considering the perturbation of $\hat{H}_\textrm{rotor}+\hat{H}_\textrm{light}+\hat{H}_\textrm{nucl}^\mathbf{B}+\hat{H}_\textrm{rotor}^\mathbf{B}+\hat{H}^{\mathbf{B}^2}+\hat{H}_\textrm{light}^\mathbf{B}$ by $\delta\hat{H}+\delta\hat{H}_\mathbf{B}$ to first order. Our approach is to consider each distinct pair of values of $\sum_j g_j m_j=\sum_{j'}g_{j'}m'_{j'}$ and $m\in\{-J,\dots,J\}$ in turn and diagonalise the matrix with elements of the form
\begin{equation}
\left(\prod_j \langle I_j,m_j |\right) \langle J_{\tau,m}|\left(\delta\hat{H}+\delta\hat{H}^\mathbf{B}\right)|J_{\tau,m}\rangle \left(\prod_{j'}| I_{j'},m'_{j'}\rangle\right). \nonumber
\end{equation}
The energy shifts thus obtained give rise in particular to hyperfine structure in the rotational spectrum of the molecule in the presence of the light and $\mathbf{B}$. 

Again, the perturbative results described above suffice to illustrate the basic features introduced by $\mathbf{B}$ but should not be used in lieu of a numerical diagonalisation of $\hat{H}''$ in general.


\section{Calculated molecular properties}
In the present appendix we report the calculated molecular properties upon which FIG. \ref{Alaninespectrumfigure}, FIG. \ref{Isotopicspectrumfigure}, FIG. \ref{Tartaricspectrumfigure} and FIG. \ref{Ibuprofenspectrumfigure} are based.

We evaluated
\begin{eqnarray}
A &=&\frac{\hbar^2}{2\sum_j M_j ( Y_j^2+Z_j^2)}, \\
B &=&\frac{\hbar^2}{2\sum_j M_j ( Z_j^2+X_j^2)} \\
C &=&\frac{\hbar^2}{2\sum_j M_j ( X_j^2+Y_j^2)}
\end{eqnarray}
using the nuclear coordinates $X_j$, $Y_j$ and $Z_j$ tabulated below together with
\begin{center}
\begin{tabular}{ c | c } 
& mass / 10$^{-26}$ kg \\ \hline
$^{1}$H & 0.1673533 \\ 
$^{12}$C & 1.9926468 \\ 
$^{13}$C & 2.1642716 \\ 
$^{16}$O & 2.6560180 \\ 
\end{tabular}
\end{center}
for the masses $M_j$. The NWChem computational chemistry program \cite{Valiev 10, Autschbach 11} was employed to calculate the electronic energy eigenstates $|k\rangle$ and associated electronic energy eigenvalues $\hbar\omega_k$ as
\begin{equation}
\hat{H}_\textrm{elec}|k\rangle=\hbar\omega_k |k\rangle
\end{equation}
with
\begin{eqnarray}
&& \hat{H}_\textrm{elec}=\Bigg[ \sum_i\frac{\hat{P}^{i}_A\hat{P}^{i}_A}{2 m_e}  \\
&& +\sum_i \sum_{i'>i}\frac{e^2}{4\pi\epsilon_0 \sqrt{\left(\hat{X}_i-\hat{X}_{i'}\right)^2+\left(\hat{Y}_i-\hat{Y}_{i'}\right)^2+\left(\hat{Z}_i-\hat{Z}_{i'}\right)^2}} \nonumber \\
&&-\sum_i \sum_{j}\frac{\mathcal{Z}_je^2 }{4\pi\epsilon_0\sqrt{\left(\hat{X}_i-X_{j}\right)^2+\left(\hat{Y}_i-Y_{j}\right)^2+\left(\hat{Z}_i-Z_{j}\right)^2}} \Bigg] \nonumber
\end{eqnarray}
the electronic Hamiltonian \cite{Born 27, Barron 04, Bunker 05, Atkins 11}, where the $\hat{P}_A^i$ are components of the canonical linear momentum of the $i$th electron; $m_e$ is the mass of the electron; $e$ is the magnitude of the electronic charge; $\hat{X}_i$, $\hat{Y}_i$ and $\hat{Z}_i$ are the coordinates of the $i$th electron and $\mathcal{Z}_j$ is the atomic number of the $j$th nucleus. These gave \cite{Barron 04, Rosenfeld 28, Craig 98}
\begin{eqnarray}
\!\!\!\!\!\!\!\!\!\!\alpha_{XX}&=&\frac{2}{\hbar}\sum_k  \frac{\omega_{k0}}{\omega_{k0}^2-c^2|\mathbf{k}|^2}\Re(\langle 0|\hat{\mu}_X|k\rangle\langle k|\hat{\mu}_X|0\rangle), \\
\!\!\!\!\!\!\!\!\!\!G'_{XX}&=&-\frac{2}{\hbar}\sum_k  \frac{ c|\mathbf{k}|}{\omega_{k0}^2-c^2|\mathbf{k}|^2}\Im(\langle 0|\hat{\mu}_X|k\rangle\langle k|\hat{m}_X|0\rangle) \\
\!\!\!\!\!\!\!\!\!\!A_{X,YZ}&=&\frac{2}{\hbar}\sum_k  \frac{\omega_{k0}}{\omega_{k0}^2-c^2|\mathbf{k}|^2}\Re(\langle 0|\hat{\mu}_X|k\rangle\langle k|\hat{\Theta}_{YZ}|0\rangle)
\end{eqnarray}
with
\begin{eqnarray}
\mu_X&=&-e\sum_i \hat{X}_i, \\
\hat{m}_X&=&-\frac{ e }{2 m_e} \sum_i ( \hat{Y}_i\hat{P}^i_Z-\hat{Z}_i\hat{P}^i_Y) \\
\hat{\Theta}_{YZ}&=&-\frac{3 e}{2} \sum_i \hat{Y}_i \hat{Z}_i,
\end{eqnarray}
for example, where $|0\rangle$ and $\hbar\omega_0$ pertain to the ground state in particular. Then \cite{Buckingham 71, Autschbach 11} 
\begin{equation}
B_{XX}= -\frac{1}{c|\mathbf{k}|}G_{XX}'+\frac{1}{3}(A_{Y,ZX}-A_{Z,XY}), 
\end{equation}
for example. Note that the nuclei are held here in the same, rigid constellation for different electronic states with the nuclear and electronic centres of mass regarded as one and the same \cite{Flygare 74, Bunker 05}. Myriad corrections to this model, not least the inclusion of the vibrational degrees of freedom of the molecule, might be entertained in more refined calculations. We found the b3lyp exchange functionals to be more reliable for the smaller molecules here and the xcamb88 exchange functionals to be more reliable for the larger ones.

For the lowest energy conformer of (S)-propylene glycol (upper signs) or (R)-propylene glycol (lower signs)
\begin{center}
\begin{tabular}{ c | c|  c| c } 
 & $X$ / 10$^{-10}$ m & $Y$ / 10$^{-10}$ m & $Z$ / 10$^{-10}$ m \\ \hline
$^{1}$H & $\pm$0.4466147 & $\pm$1.5810350 & $\mp$0.1435728 \\
$^{1}$H & $\mp$1.9321676 & $\mp$0.6877813 & $\mp$1.1311508 \\
$^{1}$H & $\mp$1.9116470 & $\mp$1.6693112 & $\pm$0.3571247 \\
$^{1}$H & $\mp$2.6617996 & $\mp$0.0589727 & $\pm$0.3561426 \\
$^{1}$H & $\pm$2.1035800 & $\mp$0.0952870 & $\pm$0.9725317 \\
$^{1}$H & $\pm$0.6162274 & $\mp$0.8246294 & $\mp$1.3103970 \\
$^{1}$H & $\pm$0.7638175 & $\mp$1.7837411 & $\pm$0.1825227 \\
$^{1}$H & $\mp$0.4158017 & $\mp$0.0155709 & $\pm$1.4521401 \\
$^{12}$C & $\pm$0.7023718 & $\mp$0.7621770 & $\mp$0.2208658 \\
$^{12}$C & $\mp$0.5050277 & $\mp$0.0129932 & $\pm$0.3473245 \\
$^{12}$C & $\mp$1.8316458 & $\mp$0.6509643 & $\mp$0.0407225 \\
$^{16}$O & $\mp$0.4927279 & $\pm$1.3228057 & $\mp$0.1347442 \\
$^{16}$O & $\pm$1.9073152 & $\mp$0.0289141 & $\pm$0.0240884 \\
\end{tabular}
\end{center}
from \cite{CCC BDB}. These gave
\begin{center}
\begin{tabular}{  c | c  } 
 & value / 10$^9$\textrm{s}$^{-1}$ \\ \hline
$A / 2 \pi \hbar$ & 8.6344378 \\ 
$B / 2\pi \hbar $ & 3.5979360 \\ 
$C / 2\pi \hbar$ & 2.7849088 \\ 
\end{tabular}
\end{center}
as well as
\begin{center}
\begin{tabular}{ c | c } 
& value / 10$^{-40}$ kg$^{-1}$.s$^4$.A$^2$ \\ \hline
$\alpha_{XX}/2$ & 4.743743 \\ 
$\alpha_{YY}/2$ & 4.272322 \\ 
$\alpha_{ZZ}/2$ & 4.001518 \\ 
$|\mathbf{k}| B_{XX}$ & $\pm$0.000071 \\
$|\mathbf{k}| B_{YY}$ & $\mp$0.000043 \\ 
$|\mathbf{k}| B_{ZZ}$ & $\pm$0.000043 \\ 
\end{tabular}
\end{center}
at $2\pi/|\mathbf{k}|=5.320000\times 10^{-7}$m using DFT with the aug-cc-pVDZ basis set and the b3lyp exchange functionals.

For isotopically chiral housane, with the upper and lower signs referring to the enantiomers obtained by replacing the usual C atom at the bottom-left or bottom-right of the `house' with a $^{13}$C atom,
\begin{center}
\begin{tabular}{ c | c | c | c  } 
& $X$ / 10$^{-10}$ m & $Y$ / 10$^{-10}$ m & $Z$ / 10$^{-10}$ m \\ \hline
$^1$H & $\mp$0.7627098 & $\mp$1.4548684 & $\pm$1.1772840 \\ 
$^1$H & $\mp$0.8466154 & $\pm$1.4430824 & $\pm$1.1631991 \\ 
$^1$H & $\pm$1.1752612 & $\mp$1.2399820 & $\mp$1.1060364 \\ 
$^1$H & $\pm$1.1011834 & $\pm$1.3185155 & $\mp$1.1184709 \\
$^1$H & $\pm$1.7782637 & $\mp$1.1284728 & $\pm$0.5631238 \\  
$^1$H & $\pm$1.7091580 & $\pm$1.2582981 & $\pm$0.5515232 \\
$^1$H & $\mp$1.1577942 & $\mp$0.0294577 & $\mp$1.5745936 \\ 
$^1$H & $\mp$2.3794033 & $\mp$0.0582261 & $\mp$0.2161851 \\
$^{12}$C & $\mp$0.3940307 & $\mp$0.7640611 & $\pm$0.4240163 \\
$^{12}$C & $\mp$0.4383745 & $\pm$0.7674796 & $\pm$0.4165727 \\
$^{12}$C & $\pm$0.9877190 & $\pm$0.8229423 & $\mp$0.1461194 \\
$^{12}$C & $\mp$1.3291769 & $\mp$0.0291823 & $\mp$0.4969827 \\ 
$^{13}$C & $\pm$1.0330404 & $\mp$0.7423828 & $\mp$0.1385121 \\ 
\end{tabular}
\end{center}
\normalsize
from \cite{CCC BDB}. These gave
\begin{center}
\begin{tabular}{ c | c  } 
& value / 10$^9$\textrm{s}$^{-1}$ \\ \hline
$A / 2 \pi \hbar$ & 9.0331364 \\ 
$B / 2\pi \hbar $ & 5.9997989 \\ 
$C / 2\pi \hbar$ & 4.6834429 \\ 
\end{tabular}
\end{center}
as well as
\begin{center}
\begin{tabular}{ c | c } 
 & value / 10$^{-40}$ kg$^{-1}$.s$^4$.A$^2$ \\ \hline
$\alpha_{XX}/2$ & 5.215255 \\ 
$\alpha_{YY}/2$ & 4.874422 \\ 
$\alpha_{ZZ}/2$ & 4.509241 \\ 
$|\mathbf{k}|B_{XX}$ & $\mp$0.000014 \\ 
$|\mathbf{k}|B_{YY}$ & $\pm$0.000012  \\ 
$|\mathbf{k}|B_{ZZ}$ & $\pm$0.000002 \\ 
\end{tabular}
\end{center}
at $2\pi/|\mathbf{k}|=5.320000\times 10^{-7}$m using DFT with the aug-cc-pVDZ basis set and the b3lyp exchange functionals.

For mesotartaric acid
\begin{center}
\begin{tabular}{ c | c | c | c } 
 & $X$ / 10$^{-10}$ m & $Y$ / 10$^{-10}$ m & $Z$ / 10$^{-10}$ m \\ \hline
$^{1}$H & 3.2759269 & -1.4121684 & 0.0008288 \\ 
$^{1}$H & 0.3010403 & 0.0938862 & 1.4929986 \\
$^{1}$H & 0.4320884 & 1.9990179 & -0.6188789 \\
$^{1}$H & -0.3010403 & -0.0938862 & -1.4929986 \\ 
$^{1}$H & -0.4320884 & -1.9990184 & 0.6188789 \\
$^{1}$H & -3.2759269 & 1.4121684 & -0.0008288 \\
$^{12}$C & 1.9010044 & 0.0508956 & 0.0829577 \\ 
$^{12}$C & 0.4800303  & 0.3819448 & 0.4570136 \\
$^{12}$C & -0.4800303 & -0.3819448 & -0.4570136 \\
$^{12}$C & -1.9010044 & -0.0508956 & -0.0829577 \\ 
$^{16}$O & 0.2630994 & 1.7859661 & 0.3091107 \\
$^{16}$O & -0.2630994 & -1.7859661  & -0.3091107 \\
$^{16}$O & -2.6250383 & -0.910888 & 0.3600024 \\
$^{16}$O & -2.3629433 & 1.2001377 & -0.2408646 \\ 
$^{16}$O & 2.3629433 & -1.2001377 & 0.2408646 \\
$^{16}$O & 2.6250383 & 0.91088792 & -0.3600024 \\ 
\end{tabular}
\end{center}
from \cite{PDBE}. These gave
\begin{center}
\begin{tabular}{ c | c } 
& value  / 10$^9$\textrm{s}$^{-1}$ \\ \hline
$A / 2 \pi \hbar$ & 2.4087195 \\ 
$B / 2\pi \hbar $ & 0.9442462 \\ 
$C / 2\pi \hbar$ & 0.7158046 \\ 
\end{tabular}
\end{center}
as well as
\begin{center}
\begin{tabular}{ c | c } 
& value / 10$^{-40}$ kg$^{-1}$.s$^4$.A$^2$ \\ \hline
$\alpha_{XX}$/2 & 7.107610 \\ 
$\alpha_{YY}$/2 & 6.572130 \\ 
$\alpha_{ZZ}$/2 & 4.668277 \\ 
\end{tabular}
\end{center}
at $2\pi/|\mathbf{k}|=5.320000\times 10^{-7}$m using DFT with the aug-cc-pVDZ basis set and the xcamb88 exchange functionals.

For \textsc{L}-tartaric acid (upper signs) or \textsc{D}-tartaric acid (lower signs)
\begin{center}
\begin{tabular}{ c | c | c | c } 
& $X$ / 10$^{-10}$ m & $Y$ / 10$^{-10}$ m & $Z$ / 10$^{-10}$ m \\ \hline
$^1$H & $\mp$3.2541887 & $\mp$1.2335262 & $\mp$0.9939221 \\ 
$^1$H & $\mp$0.3058380 & $\mp$1.1187253 & $\pm$1.1996687 \\ 
$^1$H & $\mp$0.4158265 & $\pm$1.7004711 & $\pm$0.8191299 \\ 
$^1$H & $\pm$0.3057809 & $\mp$1.1165122 & $\mp$1.2014264 \\ 
$^1$H & $\pm$0.4159136 & $\pm$1.7022874 & $\mp$0.8160007 \\ 
$^1$H & $\pm$3.2541246 & $\mp$1.2347113 & $\pm$0.9920453 \\ 
$^{12}$C & $\mp$1.9009903 & $\mp$0.2271198 & $\pm$0.0963777 \\ 
$^{12}$C & $\mp$0.4789110 & $\mp$0.2284118 & $\pm$0.5961522 \\ 
$^{12}$C & $\pm$0.4788990 & $\mp$0.2268267 & $\mp$0.5959982 \\ 
$^{12}$C & $\pm$1.9009784 & $\mp$0.2271187 & $\mp$0.0972238 \\ 
$^{16}$O & $\pm$0.2528037 & $\pm$0.9395841 & $\mp$1.3883655 \\ 
$^{16}$O & $\mp$0.2527556 & $\pm$0.9371754 & $\pm$1.3897129 \\ 
$^{16}$O & $\mp$2.6439203 & $\pm$0.6777474 & $\pm$0.3909590 \\ 
$^{16}$O & $\mp$2.3421375 & $\mp$1.2337139 & $\mp$0.6740672 \\ 
$^{16}$O & $\pm$2.6439554 & $\pm$0.6780121 & $\mp$0.3898775 \\ 
$^{16}$O & $\pm$2.3420745 & $\mp$1.2345235 & $\pm$0.6721894 \\ 
\end{tabular}
\end{center}
from \cite{PDBE}. These gave
\begin{center}
\begin{tabular}{ c | c } 
& value / 10$^9$\textrm{s}$^{-1}$ \\ \hline
$A / 2 \pi \hbar$ & 2.5070907 \\ 
$B / 2\pi \hbar $ & 0.8266659 \\ 
$C / 2\pi \hbar$ & 0.8141349 \\ 
\end{tabular}
\end{center}
as well as
\begin{center}
\begin{tabular}{ c | c } 
& value / 10$^{-40}$ kg$^{-1}$.s$^4$.A$^2$ \\ \hline
$\alpha_{XX}$/2 & 7.047660 \\ 
$\alpha_{YY}$/2 & 5.985290 \\ 
$\alpha_{ZZ}$/2 & 5.183870 \\ 
$|\mathbf{k}|B_{XX}$ & $\mp$0.000135 \\ 
$|\mathbf{k}|B_{YY}$ & $\mp$0.000061 \\ 
$|\mathbf{k}|B_{ZZ}$ & $\pm$0.000106 \\ 
\end{tabular}
\end{center}
at $2\pi/|\mathbf{k}|=5.320000\times 10^{-7}$m using DFT with the aug-cc-pVDZ basis set and the xcamb88 exchange functionals.

For a particular conformer of (S)-ibuprofen (upper signs) or (R)-ibuprofen (lower signs)
\begin{center}
\begin{tabular}{ c | c | c | c } 
& $X$ / 10$^{-10}$ m & $Y$ / 10$^{-10}$ m & $Z$ / 10$^{-10}$ m \\ \hline
$^1$H & $\mp$3.5546960 & $\mp$0.4784029 & $\pm$1.4387987 \\ 
$^1$H & $\mp$3.4747945 & $\pm$1.6950577 & $\pm$0.1921546 \\ 
$^1$H & $\mp$3.3823075 & $\pm$0.8486281 & $\mp$1.3324531 \\ 
$^1$H & $\pm$2.9373284 & $\pm$1.0865311 & $\pm$1.1234046 \\ 
$^1$H & $\mp$3.9443128 & $\mp$2.6040280 & $\pm$0.4911568 \\ 
$^1$H & $\mp$3.4724363 & $\mp$1.7546121 & $\mp$1.3484102 \\ 
$^1$H & $\mp$2.3054259 & $\mp$1.9918137 & $\mp$0.0270300 \\ 
$^1$H & $\mp$5.6657649 & $\pm$0.6201705 & $\pm$0.7079084 \\ 
$^1$H & $\mp$5.6258789 & $\mp$0.2858348 & $\mp$0.8155477 \\ 
$^1$H & $\mp$5.8324187 & $\mp$1.1402660 & $\pm$0.7236792 \\ 
$^1$H & $\mp$1.5742011 & $\pm$1.7750365 & $\pm$1.7326724 \\ 
$^1$H & $\mp$1.3374885 & $\mp$0.3265641 & $\mp$2.0207837 \\ 
$^1$H & $\pm$0.8739307 & $\pm$1.7071898 & $\pm$1.9287022 \\ 
$^1$H & $\pm$1.1055906 & $\mp$0.4041921 & $\mp$1.8304289 \\ 
$^1$H & $\pm$3.2627482 & $\pm$0.8187643 & $\mp$1.9162982 \\ 
$^1$H & $\pm$2.9931749 & $\pm$2.3568962 & $\mp$1.0761332 \\ 
$^1$H & $\pm$4.4542421 & $\pm$1.4111836 & $\mp$0.7560341 \\ 
$^1$H & $\pm$4.8645648 & $\mp$1.8010574 & $\pm$0.4902054 \\ 
$^{12}$C & $\mp$3.8003866 & $\mp$0.4464582 & $\pm$0.3694008 \\ 
$^{12}$C & $\mp$3.1054889 & $\pm$0.7708095 & $\mp$0.2722255 \\ 
$^{12}$C & $\mp$1.6001147 & $\pm$0.7272390 & $\mp$0.1538826 \\ 
$^{12}$C & $\pm$1.1604943 & $\pm$0.6460930 & $\pm$0.0635358 \\ 
$^{12}$C & $\pm$2.6463667 & $\pm$0.6027347 & $\pm$0.1805871 \\ 
$^{12}$C & $\mp$3.3505376 & $\mp$1.7674263 & $\mp$0.2600192 \\ 
$^{12}$C & $\mp$5.3188572 & $\mp$0.3050291 & $\pm$0.2357283 \\ 
$^{12}$C & $\mp$0.9900398 & $\pm$1.2986742 & $\pm$0.9507160 \\ 
$^{12}$C & $\mp$0.8589403 & $\pm$0.1160528 & $\mp$1.1520557 \\ 
$^{12}$C & $\pm$0.4000166 & $\pm$1.2577947 & $\pm$1.0601195 \\ 
$^{12}$C & $\pm$0.5309162 & $\pm$0.0751708 & $\mp$1.0424522 \\ 
$^{12}$C & $\pm$3.3803301 & $\pm$1.3387787 & $\mp$0.9590368 \\ 
$^{12}$C & $\pm$3.1879902 & $\mp$0.8277889 & $\pm$0.2802998 \\ 
$^{16}$O & $\pm$4.5386050 & $\mp$0.8782413 & $\pm$0.4247431 \\ 
$^{16}$O & $\pm$2.4918258 & $\mp$1.8341471 & $\pm$0.2444575 \\ 
\end{tabular}
\end{center}
which gave
\begin{center}
\begin{tabular}{ c | c } 
& value / 10$^9$\textrm{s}$^{-1}$ \\ \hline
$A / 2 \pi \hbar$ & 1.4836697 \\ 
$B / 2\pi \hbar $ & 0.2581787 \\ 
$C / 2\pi \hbar$ & 0.2415111 \\ 
\end{tabular}
\end{center}
as well as
\begin{center}
\begin{tabular}{ c | c } 
& value / 10$^{-40}$ kg$^{-1}$.s$^4$.A$^2$ \\ \hline
$\alpha_{XX}/2$ & 17.949840 \\ 
$\alpha_{YY}/2$ & 11.455125 \\ 
$\alpha_{ZZ}/2$ & 11.200545 \\ 
$|\mathbf{k}|B_{XX}$ & $\pm$0.000484 \\ 
$|\mathbf{k}|B_{YY}$ & $\mp$0.000338  \\ 
$|\mathbf{k}|B_{ZZ}$ & $\mp$0.000210 \\ 
\end{tabular}
\end{center}
at $2\pi/|\mathbf{k}|=5.320000\times 10^{-7}$m using DFT with the 6-311+G$^\ast$ basis set and the xcamb88 exchange functionals.


\section{Functionality of the chiral rotational spectrometer}
\label{functionality}
In the present appendix we give a quantitative model of the chiral rotational spectrometer discussed in \S\ref{Chiralrotationalspectrometer}. Our derivation borrows heavily from \cite{McGurk 74 b, Balle 80, Balle 81, Campbell 81 a, Campbell 81 b}.

\begin{figure}[h!]
\centering
\includegraphics[width=0.9\linewidth]{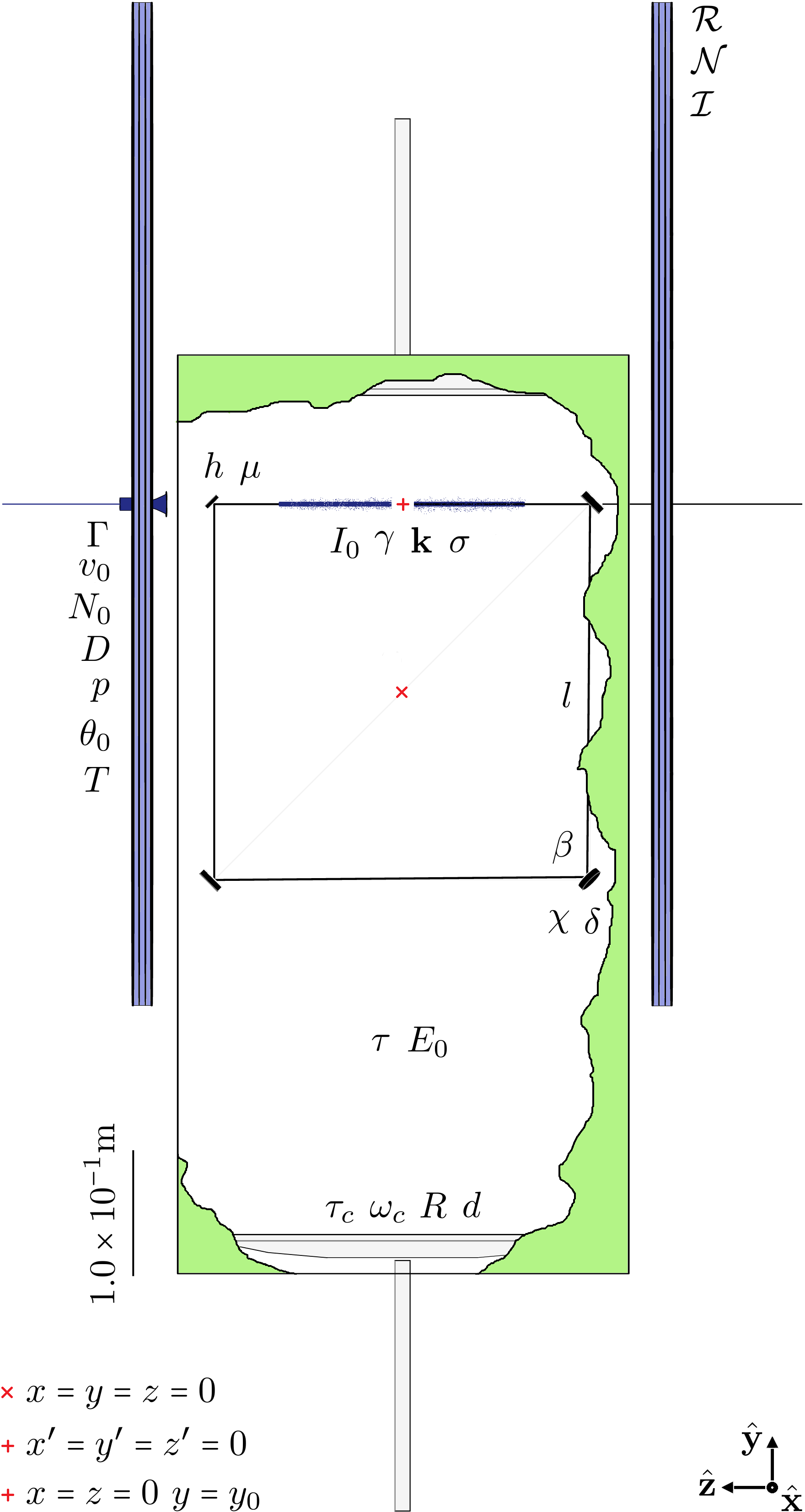}\\
\caption{\small Some of the parameters of our model and an indication of their significance, for quick reference: $\Gamma$, $v_0$, $N_0$, $D$, $p$, $\theta_0$ and $T$ pertain to the pulsed supersonic expansion nozzle and the molecular pulses; $l$, $\beta$, $\chi$ and $\delta$ pertain to the optical cavity; $I_0$, $\gamma$, $\mathbf{k}$ and $\sigma$ pertain to the light in the active region; $\tau_c$, $\omega_c$, $R$ and $d$ pertain to the microwave cavity; $\tau$ and $E_0$ pertain to the polarising microwave pulses and $\mathcal{R}$, $\mathcal{N}$ and $\mathcal{I}$ pertain to Helmholtz coils which produce the static magnetic field, as described in the text.} 
\label{Ann}
\end{figure}

Let us focus our attention here upon a particular design in which the molecular pulses are assumed to have the usual form but with a sharp angular collimation, the optical cavity is of the skew-square ring variety, the microwave cavity is of the Fabry-P\'{e}rot variety with spherical mirrors and the static magnetic field is produced by Helmholtz coils, as seen in FIG. \ref{CRSprot} and also FIG. \ref{Ann}. We consider a single form of molecule, taking $t=0$ to coincide with the onset of a polarising microwave pulse for a particular measurement. The spectrum for a mixture might then be taken as the sum of the spectra attributable to the different molecular forms present. We place the origin of $x$, $y$, $z$ at the centre of the microwave cavity, with the $y$ axis parallel to the axis of the microwave cavity and the direction of propagation of the light in the active region defining the $+z$ direction. In addition we introduce a secondary set of axes $x'$, $y'$, $z'$ which are parallel to $x$, $y$, $z$ but have their origin at the centre of the active region, located at $\mathbf{r}_0=y_0\hat{\mathbf{y}}$ with respect to $x$, $y$, $z$. The Helmholtz coils are centred upon $\mathbf{r}_0$ where they produce a static magnetic field in the $z$ direction. We imagine perfect vacuum save, of course, for the molecules themselves and any atoms that accompany them. Stray fields and radiation, including the earth's gravitational and magnetic fields and background blackbody radiation, are neglected, as is noise. A number of possible interactions, including the formation of clusters and other such complications within the molecular pulses, adhesion of the molecules to the light mirrors, changes in the resonant frequencies of the optical cavity due to heating by the light or refraction by the molecules and perturbation of the operation of the microwave cavity by the light mirrors, are neglected. These may need to be considered more carefully in some circumstances. 

Our model should be well suited to values such as \cite{RingBook, Bilger 90, McGurk 74 a, McGurk 74 b, McGurk 74 c, Balle 79, Balle 80, Balle 81, Campbell 81 a, Campbell 81 b, Legon 83, Harmony 95, Suenram 99, Brown 03}
\begin{eqnarray}
y_0&\in&[l-d/2,d/2], \nonumber \\
I_0&=& 1.00\times 10^{11}\textrm{kg.s}^{-3}, \nonumber \\
\gamma &=& 5.64\times 10^{-4}\textrm{m}, \nonumber \\
2\pi/|\mathbf{k}|&=& 5.320000\times10^{-7}\textrm{m}, \nonumber \\
|\sigma|&=&1.000, \nonumber \\
l&=&3.000000\times 10^{-1}\textrm{m}, \nonumber \\
|\beta|&=&1.000\times10^{-1}, \nonumber \\
\chi&=&4.0\times 10^{-3}, \nonumber \\
\delta&=&1\times10^{-4}, \nonumber \\
\mathcal{R}&=& 4.000\times 10^{-1}\textrm{m}, \nonumber \\
8\mu_0 \mathcal{N}|\mathcal{I}|/5\sqrt{5}\mathcal{R}&\in& [1.000\times 10^{-1}\textrm{kg.s}^{-2}\textrm{.A}^{-1}, \nonumber \\
&& 1.000\times 10^0\textrm{kg.s}^{-2}\textrm{.A}^{-1}], \nonumber \\
\Gamma/\pi&=& 5.0\times 10^3\textrm{s}^{-1}, \nonumber \\
v_0&=& 10^2\textrm{m.s}^{-2}, \nonumber \\
h&=& 2.00\times10^{-1}\textrm{m}, \nonumber \\
\mu&=& 1.0\times 10^{-2}, \nonumber \\
N_0&=&10^{23}\textrm{m}^{-3}, \nonumber \\
D&=& 1\times10^{-3}\textrm{m}, \nonumber \\
p&\in& [-5\times10^{-1},3\times10^0],  \nonumber    \\
\theta_0&=& 1.0\times10^{-3}, \nonumber \\
T&=&10^0\textrm{K}, \nonumber \\
\tau&=& 10^{-6}\textrm{s}, \nonumber \\
E_0 \mu_z^{\beta\alpha} \tau  /\hbar &\lesssim& \pi/2, \nonumber \\
\tau_c &=& 10^{-7}\textrm{s}, \nonumber \\
\omega_c/2\pi&\in&[4.50000\times10^9\textrm{s}^{-1},1.80000\times10^{10}\textrm{s}^{-1}], \nonumber \\
R&=&8.40000\times10^{-1}\textrm{m} \nonumber \\
d&\in&[5.00000\times10^{-1}\textrm{m},7.00000\times10^{-1}\textrm{m}], \nonumber 
\end{eqnarray}
for example, with the symbols as defined below and indicated in FIG. \ref{Ann}. Note in particular the implied circulating light power of $1.00\times10^5$kg.m$^2$.s$^{-3}$ in the optical cavity, which should be achievable using an input light power of $1.00\times10^2$kg.m$^2$.s$^{-3}$ or less, assuming a transmittance of $2.50\times10^{-4}$ or less for each light mirror and neglecting loss \cite{Bilger 90, Meng 05, Gold 14, F13}. Significantly higher circulating light powers than this have certainly been demonstrated, also in the context of molecular alignment \cite{Deppe 15}.

Let $\hat{H}''(\mathbf{r})$ be the effective Hamiltonian describing the rotational and nuclear-spin degrees of freedom of a molecule the centre of mass of which is notionally held fixed at some position $\mathbf{r}=x\hat{\mathbf{x}}+y\hat{\mathbf{y}}+z\hat{\mathbf{z}}$ in the active region. The energy eigenstates $|r(\mathbf{r})\rangle$ and associated energy eigenvalues $\hbar \omega_r(\mathbf{r})$ of $\hat{H}''(\mathbf{r})$ satisfy 
\begin{equation}
\hat{H}''(\mathbf{r})|r (\mathbf{r})\rangle=\hbar\omega_r (\mathbf{r})|r(\mathbf{r})\rangle.
\end{equation}
It is convenient to partition these as
\begin{eqnarray}
|r(\mathbf{r})\rangle &=& |r\rangle +\Delta |r(\mathbf{r})\rangle \\
\hbar\omega_r(\mathbf{r}) &=&\hbar\omega_r +\hbar\Delta \omega_r(\mathbf{r})
\end{eqnarray}
with $|r\rangle$ and $\hbar\omega_r$ denoting the particular forms taken at $\mathbf{r}=\mathbf{r}_0$ in the absence of the light. We assume that the $\Delta |r(\mathbf{r})\rangle$ constitute but small corrections to the $|r\rangle$ of interest, as, by construction throughout the active region, the static magnetic field is highly uniform and directed essentially parallel to the direction of propagation of the light, so that essentially the same quantisation axis for the rotational degrees of freedom of the molecules is favoured by both. We therefore neglect the $\Delta |r(\mathbf{r})\rangle$ and take $|r(\mathbf{r})\rangle=|r\rangle$ in what follows. We also assume that the $\hbar \Delta\omega_r(\mathbf{r})$ constitute but small corrections to the $\hbar\omega_r$ of interest. We nevertheless retain the $\hbar\Delta\omega_r(\mathbf{r})$ in what follows unless otherwise stated, as they appear in the arguments of sensitive mathematical functions and are the essence of chiral rotational spectroscopy. The following explicit forms might be employed:
\begin{eqnarray}
\hat{H}''(\mathbf{r}) & = & \frac{1}{\hbar^2}\left(A\hat{J}_X^2+B\hat{J}_Y^2+C\hat{J}_Z^2\right) \label{CHERN} \nonumber \\*
&-& \frac{I(\mathbf{r})}{2\epsilon_0 c} \Big\{ \Big[ \left(\cos^2\eta\cos^2\iota+\sin^2\eta\sin^2\iota\right)\hat{\ell}_{xA}\hat{\ell}_{xB}  \nonumber \\*
&& +\cos2\eta\sin2\iota \hat{\ell}_{xA}\hat{\ell}_{yB} \nonumber \\*
&& +\left(\cos^2\eta\sin^2\iota+\sin^2\eta\cos^2\iota\right)\hat{\ell}_{yA}\hat{\ell}_{yB}\Big]\alpha_{AB} \nonumber \\*
&& +\sigma|\mathbf{k}|\left(\hat{\ell}_{xA}\hat{\ell}_{xB}+\hat{\ell}_{yA}\hat{\ell}_{yB}\right)B_{AB} \Big\} \nonumber \\*
& - & \frac{\mu_N}{\hbar} \sum_j  g_j \hat{I}_z^j B_z(\mathbf{r}) \nonumber \\*
& - & \frac{\mu_N}{2\hbar}\left(\hat{\ell}_{zA} \hat{J}_B +\hat{J}_B\hat{\ell}_{zA}\right)g_{AB} B_z(\mathbf{r}) \nonumber \\*
& - & \frac{1}{2}\hat{\ell}_{zA}\hat{\ell}_{zB}\chi_{AB}B_z^2(\mathbf{r}) \\*
& - & \frac{I(\mathbf{r})}{2\epsilon_0 c} \sigma   \hat{\ell}_{xA}\hat{\ell}_{yB}\hat{\ell}_{zC} \alpha'_{AB,C} B_z(\mathbf{r})  \nonumber \\*
& - & \delta\hat{H} \nonumber \\*
& - & \delta\hat{H}^\mathbf{B}(\mathbf{r}) \nonumber
\end{eqnarray}
is the effective Hamiltonian indicated in (\ref{Franceisgreat}) but here varying through the active region as a function of $\mathbf{r}$ with a more general pure polarisation state for the light, where $I(\mathbf{r})$ is the intensity profile of the light, $\mathbf{k}=|\mathbf{k}|\hat{\mathbf{z}}$ is the central wavevector of the light, $-\eta$ is the ellipticity angle of the light as in \cite{Barron 04} (with $\sigma=\sin2\eta$ here), $-\iota$ is the polarisation azimuth of the light as in \cite{Barron 04} and $B_z(\mathbf{r})$ is the $z$ component of the static magnetic field $\mathbf{B}(\mathbf{r})$, assuming the light to propagate in a fundamental mode that is not tightly focussed. This might be augmented with
\begin{equation}
I(\mathbf{r}')=I_0\exp\left(-\frac{x'^2+y'^2}{\gamma^2}\right), \label{keeper} \\
\end{equation}
which is a Gaussian transverse intensity profile, with $\gamma$ the $1/e$ width. The precise resonance frequencies $c |\mathbf{k}|/2\pi$ of the remaining longitudinal modes supported by the optical cavity depend, of course, upon the precise length $l$ of each side of the ring as well as the fold angle $\beta$, as
\begin{eqnarray}
\frac{c|\mathbf{k}|}{2\pi}= \frac{c}{4 l}\left( g+1 \pm \frac{\sqrt{2} \beta }{\pi}\right), \label{frequencies}
\end{eqnarray}
say, with $g$ the longitudinal mode number and where the upper and lower signs distinguish opposite circular polarisations \cite{RingBook, Bilger 90}. Tacit in (\ref{frequencies}) is the assumption that $1>>|\beta|>>|\chi|,\delta$, with $\chi$ and $\delta$ reflection anisotropies of the light mirrors as in \cite{Bilger 90}. Furthermore, we might take
\begin{eqnarray}
&&B_z(\mathbf{r}')=\frac{\mu_0 \mathcal{R}\mathcal{N} \mathcal{I}}{4 \pi} \int_0^{2\pi}\Bigg\{  \\
&&+\frac{\left(\mathcal{R}-x'\cos\phi-y'\sin\phi\right)}{\left[(x'-\mathcal{R}\cos\phi)^2+(y'-\mathcal{R}\sin\phi)^2+(z'-\frac{\mathcal{R}}{2})^2\right]^{3/2}} \nonumber \\
&&+\frac{\left(\mathcal{R}-x'\cos\phi-y'\sin\phi\right)}{\left[(x'-\mathcal{R}\cos\phi)^2+(y'-\mathcal{R}\sin\phi)^2+(z'+\frac{\mathcal{R}}{2})^2\right]^{3/2}}\Bigg\} \, \textrm{d}\phi, \nonumber 
\end{eqnarray}
which is the usual expression for the $z$ component of the static magnetic field produced by Helmholtz coils, with $\mathcal{R}$ and $\mathcal{N}$ the radius and number of turns of each coil and $\mathcal{I}$ the current running through each turn. Our neglect of the $x$ and $y$ components of the light's wavevectors and $\mathbf{B}(\mathbf{r})$ in extrapolating the spatially-dependent form (\ref{CHERN}) from the more idealised form (\ref{Franceisgreat}) should introduce little error as, again by construction throughout the active region, the light propagates in a near-planar fashion and $\mathbf{B}(\mathbf{r})$ is well directed. Similarly for our neglect of diffraction in (\ref{keeper}). 

Let us assume now that the energies of molecules in motion follow the forms $\hbar\omega_r(\mathbf{r})$ adiabatically and take the master equation describing the molecules to be
\begin{equation}
        \begin{split}
                \textrm{i}\hbar & \left( \frac{\partial}{\partial t}+\mathbf{v}\cdot\boldnabla \right) \sigma^{rs} (\mathbf{r},\mathbf{v},t) \label{masterequation} = \hbar\omega_{rs}(\mathbf{r}) \sigma^{rs}(\mathbf{r},\mathbf{v},t) \\
                & + \sum_w \left[ \sigma^{rw}(\mathbf{r},\mathbf{v},t) \mu^{ws}_a - \mu^{rw}_a            \sigma^{ws}(\mathbf{r},\mathbf{v},t)\right]E_a(\mathbf{r},t) \\
                & -\textrm{i}\hbar\Gamma[\sigma^{rs}(\mathbf{r},\mathbf{v},t)-\sigma^{rs}_\textrm{eq}(\mathbf{r},\mathbf{v},t)]
        \end{split}
\end{equation}
with $\mathbf{v}$ a velocity in phase space; $\boldnabla$ the gradient operator with respect to $\mathbf{r}$; the $\sigma^{rs}(\mathbf{r},\mathbf{v},t)=\langle r|\hat{\sigma}(\mathbf{r},\mathbf{v},t)|s\rangle$ matrix elements of the density operator $\hat{\sigma}(\mathbf{r},\mathbf{v},t)$; the $\hbar\omega_{rs}(\mathbf{r})=\hbar\omega_r(\mathbf{r})-\hbar\omega_s(\mathbf{r})$ energy differences; the $\boldmu^{rs}=\langle r|\hat{\boldmu}|s\rangle$ matrix elements of the single-molecule electric-dipole moment operator $\hat{\boldmu}$; $\mathbf{E}(\mathbf{r},t)$ the polarising microwave electric field; $\Gamma$ a decay rate which models decoherence due primarily to residual molecular collisions and $\sigma^{rs}_\textrm{eq}(\mathbf{r},\mathbf{v},t)$ the equilibrium density matrix \cite{Balle 81, Campbell 81 a, Campbell 81 b}. Our neglect of absorption, Raman scattering and other such processes should be justified as the molecules are only illuminated by the light for a short time and molecular collisions occur infrequently during this time. Our use of a \textit{single} decay rate (\textit{i.e.} $\Gamma\approx1/T_1\approx1/T_2$) is in accord with empirical observations \cite{McGurk 74 a, McGurk 74 b, McGurk 74 c}. Let us imagine that the molecules proceed from the nozzle in straight lines, in which case we can assign a unique velocity $\mathbf{v}(\mathbf{r})$ to each $\mathbf{r}$. Our neglect of forces including those due to light should be well justified. It is convenient then to introduce the parametrisations
\begin{eqnarray}
\sigma^{rs}(\mathbf{r},\mathbf{v},t)&=&N(\mathbf{r},t) \delta^3 [\mathbf{v}-\mathbf{v}(\mathbf{r})]\rho^{rs}(\mathbf{r},t) \label{startzoner} \\
\sigma^{rs}_\textrm{eq}(\mathbf{r},\mathbf{v},t)&=&N(\mathbf{r},t) \delta^3 [\mathbf{v}-\mathbf{v}(\mathbf{r})]\rho^{rs}_\textrm{eq}
\end{eqnarray}
with $N(\mathbf{r},t)$ the number density distribution of the molecules, $\rho^{rs}(\mathbf{r},t)$ elements of the reduced density matrix appropriate to single molecules following the trajectory defined by $\mathbf{v}(\mathbf{r})$ and $\rho^{rs}_\textrm{eq}$ elements of the reduced equilibrium density matrix appropriate to single molecules. Our failure to acknowledge the spatial variation of $\rho^{rs}_\textrm{eq}$ should be of little consequence as the $\hbar\Delta\omega_r(\mathbf{r})$ are small relative to the $\hbar\omega_r$. Let us assume further that
\begin{eqnarray}
\mathbf{v}(\mathbf{r})\cdot\boldnabla v_a (\mathbf{r})&=&0  \\
\frac{\partial N(\mathbf{r},t)}{\partial t}+\boldnabla\cdot[\mathbf{v}(\mathbf{r})N(\mathbf{r},t)]&=&0 \label{endzoner}
\end{eqnarray}
which are statements that the molecules do indeed move in straight lines and that their number is locally conserved. Integrating the master equation (\ref{masterequation}) over $\mathbf{v}$ and making use of (\ref{startzoner})-(\ref{endzoner}) we obtain the reduced master equation
\begin{equation}
        \label{RME}
        \begin{split}
                \textrm{i}\hbar & \left[\frac{\partial}{\partial t}+\mathbf{v}(\mathbf{r})\cdot\boldnabla\right] \rho^{rs}               (\mathbf{r},t) =  \hbar\omega_{rs}(\mathbf{r}) \rho^{rs}(\mathbf{r},t) \\
                & + \sum_w \left[ \rho^{rw}(\mathbf{r},t) \mu^{ws}_a - \mu^{rw}_a \rho^{ws}(\mathbf{r},t)\right]E_a(\mathbf{r},t) \\
                & -\textrm{i}\hbar\Gamma[\rho^{rs}(\mathbf{r},t)-\rho^{rs}_\textrm{eq}]
        \end{split}
\end{equation}  
which we will begin making use of shortly. The following explicit forms might be employed:
\begin{equation}
\mathbf{v}(\mathbf{r}')=v_0\frac{(x'-h\sin\mu)\hat{\mathbf{x}}+y'\hat{\mathbf{y}}+(z'-h\cos\mu)\hat{\mathbf{z}}}{\sqrt{(x'-h\sin\mu)^2+y'^2+(z'-h\cos\mu)^2}}
\end{equation}
is a velocity field describing molecules emanating radially from the nozzle orifice, with $v_0$ the speed of the molecules, $h$ the distance from $\mathbf{r}_0$ to the centre of the nozzle orifice and $\mu$ the angle from the $+z$ axis to the line joining these two points \cite{Campbell 81 b}. 
\begin{eqnarray}
N(\mathbf{r}',t) & = & \mathcal{C}(\mathbf{r}') N_0 D^2 \\*
& \times & \frac{(h- x'\sin\mu-z'\cos\mu)^{p}}{ [(x'-h\sin\mu)^2+y'^2+(z'-h\cos\mu)^2]^{1+p/2}} \nonumber
\end{eqnarray}
describes the shape of the expanding gas pulse far from the nozzle orifice as being proportional to the usual `$\cos^p\theta/r^2$' form, with $N_0$ is the number density of molecules in the nozzle reservoir, $D$ the nozzle diameter, $p$ quantifying the angular spread of a (hypothetical) freely expanding pulse \cite{Balle 80, Balle 81, Campbell 81 b} and $\mathcal{C}(\mathbf{r}')$ a modulating function included by us to account for the effects of collimation and perhaps other effects due to short pulse times and the off-centre position of the nozzle with respect to the microwave cavity mirrors. The value of $p$ is specific to the time elapsed since the opening of the nozzle and to the mixture of molecules and atoms present in the expansion, but can be regarded as being essentially constant over the course of a measurement as the shape of the molecular pulse varies relatively slowly. In one particular experiment (with no collimation) it was determined that, for example, $p=-5\times10^{-1}$ a time $3.5\times10^{-3}$s after the opening of a nozzle and $p\ge3\times10^0$ at the later time of $5.0\times10^{-3}$s, describing an initial depletion from the beam axis and increased directivity at later times, a seemingly general trend \cite{Campbell 81 b}. For a nozzle with no collimation, $\mathcal{C}(\mathbf{r}')=1$. A simple approach therefore might be to take
 \[C(\mathbf{r}')= \begin{cases} 
      1, & \cos^{-1}\left\{\frac{(h- x'\sin\mu-z'\cos\mu)}{\sqrt{(x'-h\sin\mu)^2+y'^2+(z'-h\cos\mu)^2}}\right\} \leq \theta_0 \\
      0, & \textrm{otherwise}
   \end{cases}
\]
which describes a sharp angular collimation, with half angle $\theta_0$. This disregards any effects due to collimation upon the evolution of the molecular pulse shape as embodied by $p$, but might nevertheless prove valid for $\theta_0$ suitably small, as $p$ has little effect upon $N(\mathbf{r}',t)$ close to the axis of the molecular pulse, where $`\cos^p\theta\approx 1'$.
\begin{equation}
\rho^{rs}_\textrm{eq}=\delta_{rs}\frac{\exp\left(-\frac{\hbar\omega_r}{k_B T}\right)}{\sum_{w}\exp\left(-\frac{\hbar\omega_w}{k_B T}\right)}
\end{equation}
is an equilibrium density matrix pertaining to a thermal distribution, with $T$ the temperature of the distribution. 

First, consider times $t<0$, during which there are essentially no microwaves present in the microwave cavity. Thus, we take $\mathbf{E}(\mathbf{r},t)=0$ and assume that
\begin{equation}
\rho^{rs}(\mathbf{r},t)=\rho^{rs}_\textrm{eq}
\end{equation}
which is the equilibrium solution of the reduced master equation (\ref{RME}).

Next, consider the time interval $0\le t\le \tau$, during which a microwave pulse polarises the molecules. We model the microwave pulse temporally as being of duration $\tau$ with constant amplitude $E_0$ and spatially as being in a single TEM$_{mnq}$ mode of the microwave cavity. Our neglect of the finite rise and decay times of the microwave pulse should introduce little error as the polarisation of the molecules is an integrated quantity and the microwave cavity rise / decay time $\tau_c$ is much shorter than $\tau$ \cite{Campbell 81 a, Campbell 81 b}. Thus, we take 
\begin{equation}
\mathbf{E}(\mathbf{r},t)=\frac{1}{2}\left[\tilde{\mathbf{E}}(\mathbf{r})\exp(-\textrm{i}\omega_c t)+\tilde{\mathbf{E}}^\ast(\mathbf{r})\exp(\textrm{i}\omega_c t)\right]
\end{equation}
for
\begin{equation}
\tilde{\mathbf{E}}(\mathbf{r}) = \hat{\mathbf{z}} E_0  u(\mathbf{r})\exp[\textrm{i}(\vartheta+\omega_c\tau)]
\end{equation}
with $E_0$ the microwave pulse amplitude, 
\begin{eqnarray}
u (\mathbf{r}) & = & \frac{w_0}{w(y)} H_m\left[\frac{\sqrt{2}x}{w(y)}\right]H_n\left[\frac{\sqrt{2}z}{w(y)}\right] \nonumber \\*
& \times & \exp\left[-\frac{x^2+z^2}{w^2(y)}\right] \\* 
& \times & \cos\left[\frac{\omega_c y}{c} +\frac{\omega_c (x^2+z^2)}{2R c}-\Phi(y)-\frac{\pi q}{2}\right] \nonumber 
\end{eqnarray}
the microwave mode shape,
\begin{equation}
\omega_c=\frac{\pi c}{d}\left[q+1+\frac{1}{\pi}(m+n+1)\cos^{-1}\left(1-\frac{d}{R}\right)\right] 
\end{equation}
the microwave mode angular frequency and $\vartheta$ a tunable phase angle, where
\begin{eqnarray}
w_0&=&\left[\frac{c d(2R-d)}{\omega_c}\right]^{1/2} \\
w(y)&=&w_0\left[1+\frac{2 c y}{\omega_c w_0^2}\right]^{1/2}
\end{eqnarray}
are beam waists and
\begin{equation}
\Phi(y)=\tan^{-1}\left(\frac{2 c y}{\omega_c w_0^2}\right) 
\end{equation}
is a phase factor \cite{Balle 80, Balle 81,Campbell 81 a,Campbell 81 b}. The $H_m(x)$ are Hermite polynomials, with $H_0(x)=1$; $R$ is the radius of curvature of each microwave mirror and $d$ is the separation between the microwave mirrors. Our focus upon a \textit{single} microwave mode should be acceptable as the frequency spacings between these modes are considerably larger than the frequency bandwith of the microwave cavity. Note that we have focussed our attention explicitly here upon microwaves that are linearly polarised parallel to $z$, which are appropriate for probing $\Delta m=0$ rotational transitions. The formalism is much the same for microwaves linearly polarised parallel to $x$ say, which are appropriate for probing $\Delta m=\pm 1$ rotational transitions. The microwave mode should be chosen so as to have a good overlap with the active region, of course. Let us approximate $\mathbf{v}=0$ and $\Gamma=0$ here. Our neglect of the motion and damping of the molecules should be of little consequence as the molecules move negligible distances and experience negligible damping during the course of the polarising microwave pulse. It follows then from the reduced master equation (\ref{RME}) that 
\begin{eqnarray}
\rho^{rs}(\mathbf{r},t) & = & \rho_\textrm{eq}^{rs} +  \frac{E_0}{2\textit{i}\hbar} (\rho_\textrm{eq}^{rr}-\rho_\textrm{eq}^{ss}) \mu_z^{rs} u(\mathbf{r}) \nonumber \\*
& \times & \Bigg[ \exp\left( \textrm{i}\left\{\vartheta +\frac{1}{2}[\omega_c-\omega_{rs}(\mathbf{r})]t\right\}\right) \nonumber \\*
& \quad & \times \operatorname{sinc} \left\{\frac{1}{2}\left[\omega_c-\omega_{rs}(\mathbf{r})\right]t\right\} \\*
& \quad & + \exp\left(-\textrm{i}\left\{\vartheta +\frac{1}{2}[\omega_c+\omega_{rs}(\mathbf{r})]t\right\}\right) \nonumber \\*
& \quad & \times \operatorname{sinc}\left\{\frac{1}{2}\left[\omega_c+\omega_{rs}(\mathbf{r})\right]t\right\} \Bigg] \nonumber 
\end{eqnarray}
to first order in $\mathbf{E}(\mathbf{r},t)$, with $\textrm{sinc}(x)=\sin(x)/x$. Our neglect of higher-order contributions in $\mathbf{E}(\mathbf{r},t)$ should introduce little error, assuming $E_0$ and $\tau$ to be such that the polarisation of the molecules is a little less than would result from a $\pi/2$ pulse for an analogous two-level system, say.

Finally, consider times $t>\tau$, during which the molecules exhibit a free induction decay. Taking $\mathbf{E}(\mathbf{r},t)=0$ once again, it follows from the reduced master equation (\ref{RME}) that
\begin{eqnarray}
\rho^{rs}(\mathbf{r},t) & = & \rho_\textrm{eq}^{rs}+\exp[-\Gamma(t-\tau)] \label{DECAYdensity} \\* 
& \times & \exp \left\{ -\textrm{i} \int_\tau^t \omega_{rs} [\mathbf{r}-\mathbf{v}(\mathbf{r})(t-t')]\,\textrm{d}t' \right\} \nonumber \\* 
& \times & \{\rho^{rs}[\mathbf{r}-\mathbf{v}(\mathbf{r})(t-\tau),\tau] -\rho^{rs}_\textrm{eq}\}. \nonumber
\end{eqnarray}
The polarisation
\begin{equation}
\mathbf{P}(\mathbf{r},t)=N(\mathbf{r},t) \sum_r \sum_s \boldmu^{rs} \rho^{sr}(\mathbf{r},t)  \label{LIMPBIZKITZZZ} 
\end{equation} 
associated with the molecules generates a microwave electric field $\mathbf{E}^\textrm{s}(\mathbf{r},t)$ in the microwave cavity, which we take to satisfy
\begin{equation}
        \label{Mylene1}
\left(\boldnabla^2 - \frac{1}{c^2}\frac{\partial^2}{\partial t^2}  \right)\mathbf{E}^\textrm{s}(\mathbf{r},t) = 
\frac{1}{c^2\tau_c}\frac{\partial}{\partial t} \mathbf{E}^\textrm{s}(\mathbf{r},t) +  \mu_0 \frac{\partial^2}{\partial t^2}  \mathbf{P} (\mathbf{r},t)
\end{equation}   
with $\boldnabla^2 = \boldnabla\cdot\boldnabla$ and 
\begin{equation}
\mathbf{E}^\textrm{s}(\mathbf{r},t)=\hat{\mathbf{z}}s(t)u(\mathbf{r})
\end{equation}
a single-mode form echoing that of the polarising microwave pulse \cite{Campbell 81 a}. The electric field variation $s(t)$ diminishes slowly in average magnitude over time, primarily as a result of residual molecular collisions. Let us suppose simply here that the detected lineshape $S(t)$ is $0$ for $t\le\tau$ and proportional to $s(t)$ for $t>\tau$ \cite{Campbell 81 a, Campbell 81 b}. We then calculate the Fourier transform 
\begin{equation}
\tilde{S}(\omega)= \exp(-\textrm{i}\omega\tau )\int_\tau^\infty S(t) \exp(\textrm{i}\omega t) \, \textrm{d} t.  \label{Mylene2}
\end{equation}
and regard the real part $\Re[\tilde{S}(\omega)]$ of this as being the measurement, given an appropriate choice of $\vartheta$. Using (\ref{Mylene1})-(\ref{Mylene2}) and taking
\begin{equation}
\boldnabla^2 u (\mathbf{r})= -\frac{\omega_c^2}{c^2}u(\mathbf{r})
\end{equation}
we obtain
\begin{eqnarray}
\tilde{S}(\omega) & \propto &  \omega^2 \frac{\omega_c^2-\omega^2+\frac{\textrm{i}\omega}{\tau_c}}{(\omega_c^2-\omega^2)^2+\frac{\omega^2}{\tau_c^2}} \frac{E_0}{2\textrm{i}\hbar} \label{xTinA} \nonumber \\*
& \times & \sum_r\sum_s|\mu_z^{sr}|^2 (\rho^{ss}_\textrm{eq}-\rho_\textrm{eq}^{rr}) \nonumber \\*
& \times & \int \!\!\!\!\!\int \!\!\!\!\!\int \! \! \! \! \!\int_0^\infty  N(\mathbf{r},t)   u(\mathbf{r})u[\mathbf{r}-\mathbf{v}(\mathbf{r})t] \exp[(\textrm{i}\omega-\Gamma) t] \nonumber \\*
&\times & \exp\left\{   -\textrm{i} \int_0^ t \omega_{sr}[\mathbf{r}-\mathbf{v}(\mathbf{r})(t-t')]\, \textrm{d}t'\right\} \nonumber \\*
& \times &\Bigg\{ \exp\left[ \textrm{i}\left(\vartheta +\frac{1}{2} \{\omega_c-\omega_{sr}[\mathbf{r}-\mathbf{v}(\mathbf{r})t]\}\tau\right)\right] \\* 
& \quad & \times \operatorname{sinc}\left(\frac{1}{2}\left\{\omega_c-\omega_{sr}[\mathbf{r}-\mathbf{v}(\mathbf{r})t]\right\}\tau\right) \nonumber \\*
& \quad & + \exp\left[-\textrm{i}\left(\vartheta+\frac{1}{2} \{\omega_c+\omega_{sr}[\mathbf{r}-\mathbf{v}(\mathbf{r})t]\}\tau\right)\right] \nonumber\\*
& \quad & \times \operatorname{sinc}\left(\frac{1}{2}\left\{\omega_c+\omega_{sr}[\mathbf{r}-\mathbf{v}(\mathbf{r})t]\right\}\tau\right) \Bigg\} \textrm{d}t \, \textrm{d}^3\mathbf{r} \nonumber
\end{eqnarray}
where we have made the replacement $N(\mathbf{r},t+\tau)\rightarrow N(\mathbf{r},t)$, which is justified as the shape of the molecular pulse varies slowly relative to $\tau$, and have assumed that $\boldmu^{rs}=0$ if $r=s$, which is justified as asymmetric rotors do not exhibit first-order Stark shifts. 

Let us focus our attention now upon a particular, well-isolated rotational transition. We label the nuclear-spin state manifold of the lower rotational state as $\alpha$ and the nuclear-spin state manifold of the upper rotational state as $\beta$. For $\omega\sim\omega_c\sim \omega_{\beta\alpha}\gg\Gamma$ the dominant contribution to (\ref{xTinA}) then comes when $r$ sums over $\alpha$ whilst $s$ sums over $\beta$. If we assume moreover that $|\omega_{\beta\alpha}-\omega_c|\ll|\omega_{\beta\alpha}+\omega_c|$, we can safely neglect the term proportional to $\exp(-\textrm{i}\vartheta)$. In addition, we will consider only those frequencies that lie well within the microwave cavity frequency bandwidth, as $|\omega-\omega_c|\ll1/\tau_c$, in which case we can safely make the replacement $\omega^2 (\omega_c^2-\omega^2+\textrm{i}\omega / \tau_c)/[(\omega_c^2-\omega^2)^2+ \omega^2/\tau_c^2]\rightarrow\textrm{i}Q$, with $Q=\omega_c\tau_c$ the quality factor of the microwave cavity. We are left then with
\begin{eqnarray}
\tilde{S}(\omega) & \propto & \frac{Q E_0 }{2\hbar}  \exp(\textrm{i}\vartheta) \label{generalenough} \nonumber \\*
& \times & \sum_\alpha\sum_\beta  |\mu_z^{\beta\alpha}|^2  (\rho^{\beta\beta}_\textrm{eq}-\rho_\textrm{eq}^{\alpha\alpha}) \nonumber \\*
& \times & \int \!\!\!\!\!\int \!\!\!\!\!\int \! \! \! \! \!\int_0^\infty  N(\mathbf{r},t)   u(\mathbf{r})u[\mathbf{r}-\mathbf{v}(\mathbf{r})t] \exp[(\textrm{i}\omega-\Gamma) t] \nonumber \\*
& \times &  \exp\left\{-\textrm{i} \int_0^t \omega_{\beta\alpha}[\mathbf{r}-\mathbf{v}(\mathbf{r})(t-t')]\, \textrm{d}t'\right\} \\*
& \times & \exp\left( \frac{\textrm{i}}{2} \{\omega_c-\omega_{\beta\alpha}[\mathbf{r}-\mathbf{v}(\mathbf{r})t]\}\tau \right) \nonumber \\*
& \times&\operatorname{sinc}\left(\frac{1}{2}\left\{\omega_c-\omega_{\beta\alpha}[\mathbf{r}-\mathbf{v}(\mathbf{r})t]\right\}
\tau\right) \, \textrm{d}t \, \textrm{d}^3\mathbf{r}. \nonumber
\end{eqnarray}
This form may be appreciated by considering the idealised limit
\begin{eqnarray}
\mathbf{v}(\mathbf{r})&\rightarrow&-v_0\hat{\mathbf{z}}, \nonumber \\
N(\mathbf{r}',t)&\rightarrow&n_0 \delta(x')\delta(y'), \nonumber \\
u(x_0,y_0,z+v_0 t)&\rightarrow &u(x_0,y_0,z), \nonumber \\
\omega_{\beta\alpha}(x_0,y_0,z+v_0 t)&\rightarrow&\omega_0, \nonumber \\
\exp\left( \frac{\textrm{i}}{2} \{\omega_c-\omega_{\beta\alpha}[\mathbf{r}-\mathbf{v}(\mathbf{r})t]\}\tau \right)&\rightarrow&1, \nonumber \\
\textrm{sinc}\left(\frac{1}{2}\left\{\omega_c-\omega_{\beta\alpha}[\mathbf{r}-\mathbf{v}(\mathbf{r})t]\right\}\tau\right)&\rightarrow&1 \nonumber \\
\rho_\textrm{eq}^{\beta\beta}-\rho_\textrm{eq}^{\alpha\alpha}&\rightarrow&\Delta\rho \nonumber 
\end{eqnarray}
pertaining to well-collimated molecular pulses moving parallel to the axis of the active region with $n_0$ the number of molecules per unit length on axis; motion through the microwave mode during the course of a measurement neglected; variations of $I(\mathbf{r})$ and $\mathbf{B}(\mathbf{r})$ on axis as well as hyperfine splittings neglected, leaving a unique rotational transition angular frequency $\omega_0$; the microwave cavity mode chosen to be on resonance as $\omega_c=\omega_0$ and a single value $\Delta\rho$ taken to be a fair representation of the differences in the equilibrium Boltzmann factors for the upper and lower manifolds. The rotational spectrum itself then tends towards
\begin{equation}
\Re [\tilde{S}(\omega)]\propto Q E_0 \kappa \Delta\rho n_0 \zeta  \frac{1}{\pi}\frac{\Gamma}{\Gamma^2+(\omega-\omega_0)^2}
\end{equation}
for $\vartheta=0$, with $\zeta=\pi \int u^2(x_0,y_0,z) \, \textrm{d}z /2\hbar$ a geometrical factor that accounts for the overlap between the microwave mode and the molecular beam and $\kappa=\sum_\alpha \sum_\beta |\mu_z^{\beta\alpha}|^2$ a measure of the strength of the rotational transition. This is a Lorentzian, like those plotted in \S\ref{Chiralrotationalspectra}. To approach this limit in practice would require in particular that the molecules pass through a light mirror, which might be difficult to realise without significantly compromising the optical cavity.

\begin{figure}[h!]
\centering
\includegraphics[width=\linewidth]{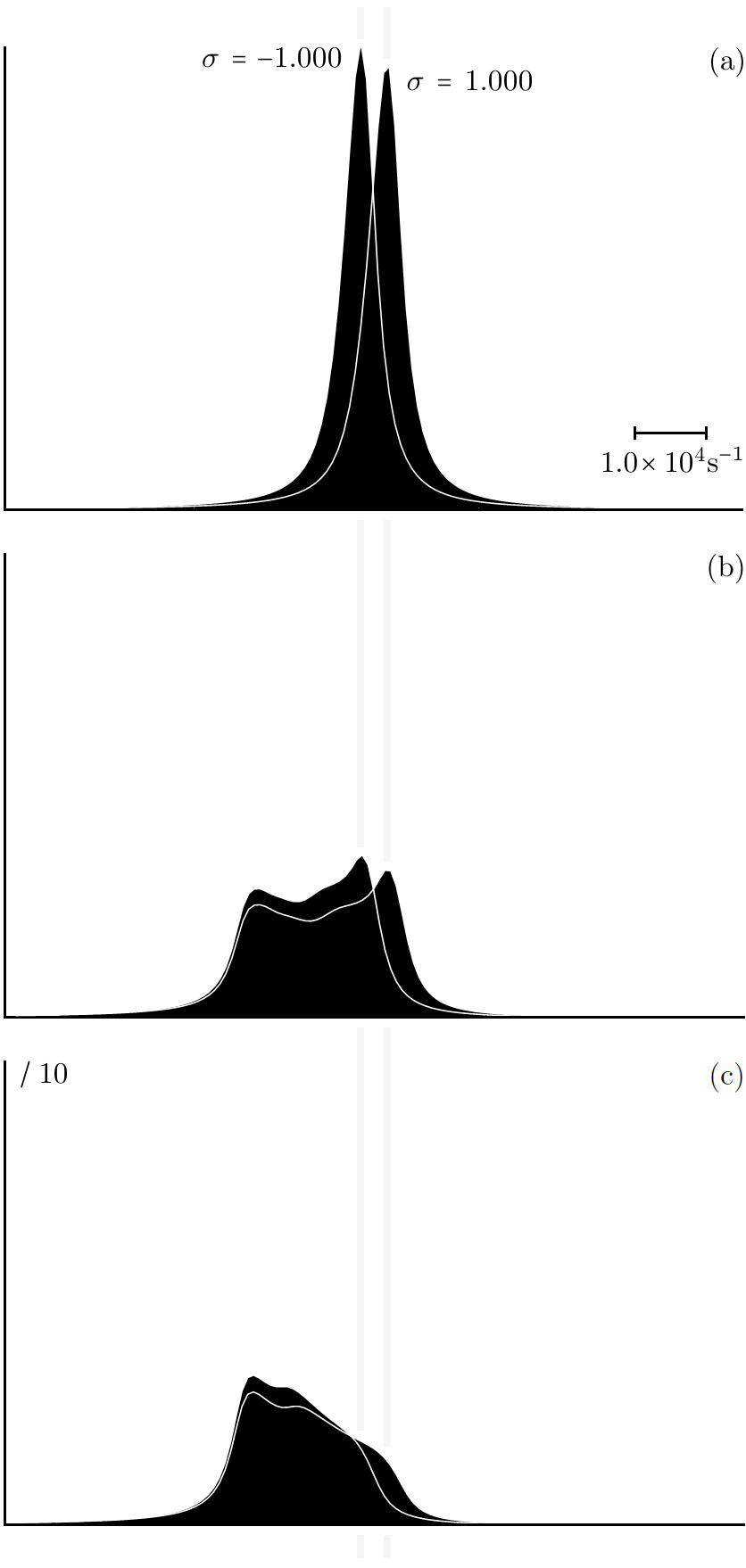} \\
\caption{\small Lineshapes for different molecular pulse geometries and opposite circular polarisations of the light.} 
\label{FullNumerical}
\end{figure}

In general (\ref{generalenough}) must be integrated numerically. Our preliminary investigations here reveal that under more realistic operating conditions the spectrometer yields a lineshape resembling a Lorentzian but with a significant broadening on one side; that closest to the microwave rotational transition frequency as it would appear in the absence of the light. This asymmetric broadening occurs because those molecules removed from the most intense regions of the light experience weaker (but not stronger) energy shifts due to the light. Chirally sensitive information can be extracted from these lineshapes in spite of their unusual forms, although for some tasks such as the determination of enantiomeric excess a fitting procedure might be required.

To illustrate these ideas let us consider $\omega_{\beta\alpha}(\mathbf{r}')/2\pi=\{1.7030023+(0.0000025+0.0000002\sigma) \exp[-(x'^2+y'^2)/\gamma^2]\}\times10^{10}\textrm{s}^{-1}$, which is representative of a reasonably magic rotational transition, be it naturally occurring or refined, for a chiral molecule with a significant but not exceptional chiroptical response, comparable to that of ibuprofen say. Depicted in FIG. \ref{FullNumerical} are fair numerical approximations to $\Re[\tilde{S}(\omega)]$ from (\ref{generalenough}) for: (a) $\mu=0.0$ and $\theta_0=1.0\times 10^{-3}$, which would require in particular that the molecules pass through a light mirror; (b) $\mu=1.0\times10^{-2}$ and $\theta_0=1.0\times 10^{-3}$, which should be viable without compromising the stability of the optical cavity; (c) $\mu=1.0\times10^{-2}$ and $\theta_0=3.0\times 10^{-3}$, which should also be viable without compromising the stability of the optical cavity but corresponds to a weaker angular collimation than is desirable. In all three cases $y_0=1.750\times 10^{-1}$m, $\gamma=5.64\times10^{-4}$m, $\sigma=\pm1.000$, $\Gamma/\pi=5.0\times10^{3}$s$^{-1}$, $v_0=1.0\times10^2$m.s$^{-1}$, $h=2.00\times10^{-1}$m, $p=2.0$, $\tau=1.0\times 10^{-6}$s, $\omega_c/2\pi=1.70301\times10^{10}$s$^{-1}$, $d=5.18838\times10^{-1}$m and $\vartheta=0.00$ in particular with operation in the TEM$_{008}$ microwave mode. The lineshapes are strongly dependent upon both the tilt ($\mu$) and angular collimation ($\theta_0$) of the molecular pulses, as we might expect. The $\sigma$-dependent splitting is nevertheless apparent in all three cases, however.

We note finally that calculating the real part of the Fourier transform $\tilde{S}(\omega)$ might not be the most transparent way of interpreting the free induction decays recorded in chiral rotational spectroscopy. A different function in which the unusual geometry inherent to the spectrometer is compensated for might be calculated instead, perhaps yielding clearer chiral rotational spectra without further work.


\section{Signal-to-noise ratio}
\label{SNRappendix}

In the present appendix we estimate the signal-to-noise ratio that might be attained for the chiral rotational spectrometer discussed in \S\ref{Chiralrotationalspectrometer} and Appendix \ref{functionality}.

The precise value of the signal-to-noise ratio will depend, of course, upon the quality of the components used and the care with which the spectrometer is built and measurements are taken, as well as the nature of the sample and the rotational line under consideration. We can nevertheless give some idea here of what should be possible by recalling that a comparable standard cavity enhanced Fourier transform microwave spectrometer gave a signal-to-noise ratio of $5\times10^{-1}$s$^{-1/2}\sqrt{\Delta t}$ with $\Delta t$ the total recording time for a measurement rate of $5\times10^0$s$^{-1}$ of the $J=2\leftarrow 1$ transition in the $^{18}$O$^{13}$C$^{34}$S molecules that exist with a natural abundance of $0.000094\%$ in a sample of $1\%$ OCS seeded in an $80:20$ Ne:He carrier gas \cite{Suenram 99}. We extrapolate from this a signal-to-noise ratio of
\begin{eqnarray}
\textrm{SNR}'&=& (9.4\times10^{-5})^{-1}( 5\times10^{-1}\textrm{s}^{-1/2}\sqrt{\Delta t}) \nonumber \\
&=& 5\times 10^3\textrm{s}^{-1/2}\sqrt{\Delta t}
\end{eqnarray}
for a \textit{pure} sample, containing a \textit{single} form of molecule ($^{18}$O$^{13}$C$^{34}$S). Our approach now is to estimate the signal-to-noise ratio $\textrm{SNR}$ of the chiral rotational spectrometer by scaling the signal-to-noise ratio $\textrm{SNR}'$ of the standard rotational spectrometer in accord with the following considerations.

\begin{enumerate}

\item[\textbf{(i)}] There will be fewer molecules in the chiral rotational spectrometer than in the standard rotational spectrometer, due in particular to the sharp angular collimation of the molecular pulses. This is perhaps the single most significant cause of signal reduction. Let us introduce
\begin{eqnarray}
f_1&=& \frac{\int_0^{\theta_0} \cos^2\theta \sin \theta \, \textrm{d}\theta}{ \int_0^{\pi/2} \cos^2\theta \sin \theta \, \textrm{d}\theta} \nonumber \\
&=& 1-\cos^3\theta_0 
\end{eqnarray}
as the ratio of the `solid angle' occupied by the molecular pulses in the chiral rotational spectrometer to the analogous quantity in the standard rotational spectrometer, with each of these `solid angles' weighted by the number density distribution of the molecular pulses, assuming $p=2$. For $\theta_0=10^{-3}$ we obtain $f_1=10^{-6}$. Further reductions in the number of molecules seem likely, for example if the molecules are of lower volatility than $^{18}$O$^{13}$C$^{34}$S and so cannot be sampled at the same density or if it not possible to prepare a pure sample due to the existence of multiple stereoisomers. Let us introduce
\begin{equation}
f_2= \frac{N_0}{ N_0'}
\end{equation}
as the ratio of the number density $N_0$ of molecules in the chiral rotational spectrometer to the analogous quantity $N_0'=10^{23}$m$^{-3}$ for the standard rotational spectrometer. We might hope that $10^{-2}\le f_2\le 10^0$ for first demonstrations of chiral rotational spectroscopy but recognise that significantly smaller values of $f_2$ may be encountered under many circumstances of practical interest.

\item[\textbf{(ii)}] The fraction of molecules in the rotational states of interest and also the coupling of the rotational transition of interest to the microwave pulses will in general differ, of course, from the particular values described above for the standard rotational spectrometer. Let us introduce
\begin{equation}
f_3=\frac{\Delta \rho}{\Delta \rho'}
\end{equation}
as the ratio of the difference $\Delta\rho$ in the equilibrium populations of the rotational states of interest in chiral rotational spectroscopy to the analogous quantity $\Delta \rho'$ for the standard rotational spectrometer. The molecules of interest in chiral rotational spectroscopy will be larger and more complicated than $^{18}$O$^{13}$C$^{34}$S, with lower rotational energies and more rotational states such that $10^{-2}\le f_3 \le 10^{-1}$, perhaps. Let us introduce
\begin{equation}
f_4=\left(\frac{\mu}{\mu'}\right)^2
\end{equation}
as the square of the ratio of the transition electric dipole-moment moment $\mu$ for the rotational states of interest in the chiral rotational spectrometer to the analogous quantity $\mu'=10^{-30}$m.s.A for the standard rotational spectrometer. The molecules of interest in chiral rotational spectroscopy will likely have larger permanent electric-dipole moments than $^{18}$O$^{13}$C$^{34}$S such that $10^0\le f_4\le 10^1$, perhaps. Let $f_5$ account for additional reductions or perhaps enhancements in signal strength such as those due to the choice of carrier gas, which can see the signal strength vary by orders of magnitude \cite{Suenram 99}. We would hope that $f_5=10^0$ under well chosen operating conditions.

\item[\textbf{(iii)}] The rate at which measurements are taken in a standard cavity enhanced Fourier transform microwave spectrometer is limited physically by the time taken to evacuate molecules between successive measurements \cite{Balle 80, Balle 81, Harmony 95, Suenram 99}. In a chiral rotational spectrometer measurements might be taken instead at an increased rate owing to the smaller number of molecules present, perhaps up to $7.5\times10^2$s$^{-1}$ \cite{Cross 82}. This is approaching continuous operation of the pulsed supersonic expansion nozzle, with measurements made around once every ten free induction decay times, assuming $\Gamma/\pi=5.0\times 10^3\textrm{s}^{-1}$ say. Let us introduce
\begin{equation}
f_6=\sqrt{\frac{\Lambda}{\Lambda'}}
\end{equation}
as the square root of the ratio of the measurement rate $\Lambda$ in the chiral rotational spectrometer to the analogous quantity $\Lambda'=5\times10^0$s$^{-1}$ in the standard rotational spectrometer. We might hope for an enhancement of $10^0 \le f_6\le\times10^1$ here.

\end{enumerate}
Finally, we take
\begin{equation}
\frac{\textrm{SNR}}{\textrm{SNR}'}=f_1 f_2 f_3 f_4 f_5 f_6.
\end{equation}
For $f_1=10^{-6}$, $10^{-2}\le f_2\le 10^0$, $10^{-2}\le f_3\le 10^{-1}$, $10^0\le f_4 \le 10^1$, $f_5=10^0$ and $10^0\le f_6\le 10^1$ we obtain $10^{-10}\le \textrm{SNR}/\textrm{SNR}'\le 10^{-5}$, or $10^{-6}\textrm{s}^{-1/2}\sqrt{\Delta t}\le \textrm{SNR}\le 10^{-1} \textrm{s}^{-1/2}\sqrt{\Delta t}$. This suggests in turn that a very agreeable chiral rotational spectrum could be obtained for a recording time of a few hours under favourable operating conditions, as discussed in \S\ref{Chiralrotationalspectrometer}.

\end{appendix}



\begin{thebibliography}{99}
\bibitem{Caveatt} We assume throughout the present paper that the value of the isotropic sum (\ref{isotropic sum}) is known.
\bibitem{F3} We neglect the possibility of spontaneous emission. This might be justified if attention is restricted to a suitably short time interval.
\bibitem{Newfootnote} For light with an electric field of the form $\Re \{ (\tilde{\alpha}\hat{\mathbf{x}}+\tilde{\beta}\hat{\mathbf{y}})\exp\left[\textrm{i}|\mathbf{k}|(z-c t)\right]\}$, say, we take $I=\epsilon_0 c (|\tilde{\alpha}|^2+|\tilde{\beta}|^2)/2$ and $\sigma=\textrm{i}(\tilde{\alpha}\tilde{\beta}^\ast-\tilde{\alpha}^\ast\tilde{\beta})/(|\tilde{\alpha}|^2+|\tilde{\beta}|^2)=\pm 1$ here.
\bibitem{F6} We neglect the possibility of absorption, Raman scattering and other such process which threaten to change the vibronic state of the molecule. This might be justified if attention is restricted to a suitably short illumination time.
\bibitem{F4} We neglect the vibrational degrees of freedom of the molecule and assume the unperturbed electronic wavefunctions to be real.
\bibitem{F8} We have assumed equal line strengths for each molecule.
\bibitem{F13} It has been shown that mirrors of the type we envisage here can sustain continuous illumination with no apparent damage by light with an intensity of at least $1.01\times10^{12}$kg.s$^{-3}$ \cite{Meng 05}
\end{thebibliography}
\end{document}